\documentclass[12pt,preprint]{aastex}

\newcommand{\gcc}{\ \mathrm{g\ cm^{-3} }}
\newcommand{\nuc}[2]{\ensuremath{\mathrm{^{#1}#2}}}
\newcommand{\FLASH}{{\sc flash}}
\begin{document} 

\title{Spontaneous Initiation of Detonations in White Dwarf Environments:
Determination of Critical Sizes}

\shorttitle{Spontaneous Initiation of Detonations}
\shortauthors{Seitenzahl et al.}

\author{
Ivo R. Seitenzahl\altaffilmark{1,2}, Casey A. Meakin\altaffilmark{2,3,4,5}, Dean M. Townsley\altaffilmark{2,4}, Don Q. Lamb\altaffilmark{3,4}, James W. Truran\altaffilmark{2,3,4,6,7}
}
\altaffiltext{1}{Department of Physics,
                 The University of Chicago,
                 Chicago, IL  60637}
\altaffiltext{2}{Joint Institute for Nuclear Astrophysics,
                 The University of Chicago,
                 Chicago, IL  60637}
\altaffiltext{3}{The Center for Astrophysical Thermonuclear Flashes,
                 The University of Chicago,
                 Chicago, IL  60637} 
\altaffiltext{4}{Department of Astronomy and Astrophysics,
                 The University of Chicago,
                 Chicago, IL  60637}	
\altaffiltext{5}{Steward Observatory,
                 The University of Arizona,
                 Tucson, AZ  85719}
\altaffiltext{6}{Enrico Fermi Institute,
                 The University of Chicago,
                 Chicago, IL  60637}
\altaffiltext{7}{Argonne National Laboratory,
                 Argonne, IL 60439}

\begin{abstract}
Some explosion models for Type Ia supernovae (SNe Ia), such as the gravitationally confined detonation (GCD) or the double detonation sub-Chandrasekhar (DDSC) models, rely on the spontaneous initiation of a detonation in the degenerate \nuc{12}{C}/\nuc{16}{O} material of a white dwarf. The length scales pertinent to the initiation of the detonation are notoriously unresolved in multi-dimensional stellar simulations, prompting the use of results of 1D simulations at higher resolution, such as the ones performed for this work, as guidelines for deciding whether or not conditions reached in the higher dimensional full star simulations successfully would lead to the onset of a detonation. Spontaneous initiation relies on the existence of a suitable gradient in self-ignition (induction) times of the fuel, which we set up with a spatially localized non-uniformity of  temperature -- a hot spot. We determine the critical (smallest) sizes of such hot spots that still marginally result in a detonation in white dwarf matter by integrating the reactive Euler equations with the hydrodynamics code \FLASH. We quantify the dependences of the critical sizes of such hot spots on composition, background temperature, peak temperature, geometry, and functional form of the temperature disturbance, many of which were hitherto largely unexplored in the literature. We discuss the implications of our results in the context of modeling of SNe Ia. \\
  
\end{abstract}

\section{Introduction}
\label{sec:intro}
Type Ia supernovae (SNe Ia) are believed to be thermonuclear explosions of accreting white dwarfs, 
powered by the energy liberated in the fusion of the initial composition to more tightly bound nuclear species, often all the way to nuclear statistical equilibrium (NSE). 
The explosion mechanism of SNe Ia is still unknown, and the large variation of their peak luminosity and spectral properties allows for a range of explosion mechanisms. 
Pure deflagration models have a lot of unprocessed carbon and oxygen at low velocities and generally turbulently mixed layers of nuclear material burned to differing degrees of completion \citep{reinecke02a,reinecke02b,gamezo03a}. This is more or less in contradiction to observations, which show a layered structure of the ejecta, and a lack of low velocity low and intermediate mass elements \citep[e.g.][]{branch04}.
Therefore, many of the currently remaining viable explosion models, such as deflagration to detonation transition (DDT) \citep[e.g.][]{khokhlov97,gamezo05,roepke07b}, gravitationally confined detonation (GCD) \citep{plewa04,townsley07,plewa07,jordan08,meakin08}, or double detonation sub-Chandrasekhar (DDSC) \citep{livne90,fink07} have detonations as a necessary ingredient. 

The spatial scales relevant for the initiation of these detonations are unresolved in even the most computationally expensive multi-dimensional simulations, and recent major efforts in the modeling of the SNe Ia explosion mechanism like \citet{fink07}, \citet{roepke07}, \citet{townsley07}, and \citet{jordan08} all rather similarly decide whether or not a detonation is launched based on the peak temperature reached above a certain density in a computational cell.
To make the call in favor of or against a successful detonation, one commonly checks whether or not a 1D reactive hydrodynamics calculation starting with a hot spot consisting of a linear temperature profile, with peak temperature equal to the hottest temperature that obtains in the simulation, leads to a detonation at that density for reasonably small values of the radius of the temperature inhomogeneity, as is done in the pioneering work of \citet{niemeyer97}. 

Such detonation conditions derived from 1D unresolved simulations, which assume a linear temperature gradient, a cold isothermal ambient medium, spherical geometry, constant density and fuel concentration, and an initially stagnant velocity flow field, are clearly an oversimplyfication of the initiation problem. 
Therefore, currently employed procedures of calling the outcome of a supernova simulation based on external detonation conditions (including the ones presented here) should be taken with a grain of salt and prudence should be exercised in their application. 
In this paper, we illustrate the uncertainties in the detonation conditions by taking a closer look at the consequences of relaxing the stringent assumptions about the temperature profile and geometry. 

Since existence of a supersonic induction time gradient due to a temperature and/or fuel concentration gradient (and to a lesser degree density) is the essential ingredient in the formation of the detonation, it is crucial that the high temperatures be reached in unburned fuel and that the gradients be not too steep. 
For unresolved temperature gradients, the spatial variation of the fuel concentration, which is an important ingredient in the determination of the induction time gradient, is unknown. 
Even when assuming the best case scenario of 100\% fuel concentration, one should allow for the possibility of different functional forms of the unresolved temperature gradient, the effects of which we explore in this paper. 
As we demonstrate in this work, for the same density, peak and background temperatures a gaussian temperature profile may not lead to a detonation where a linear profile does, even though more energy was placed into the hot spot region. 
Similarly, for some choices of environmental parameters a gaussian temperature profile leads to detonation where a linear profile does not, changing the conclusions about the success or failure to detonate of some simulations.
Additionally, as we will show, the temperature of the medium surrounding the hot spot (the background temperature) can also have a large effect on the outcome and should be taken into account as an additional parameter. 
This work furthermore includes a study of the effects of composition (different mass fractions of \nuc{16}{O}, \nuc{12}{C} and \nuc{4}{He}) on the critical sizes. 

In section 2, we give an overview about detonations, their spontaneous initiation, and dependences on environmental parameters. In section 3 we describe our method for determining critical radii for linear temperature profiles and critical decay constants for different exponential temperature profiles.  Section 4 contains a presentation of the results and trends with varying environmental parameters. Section 5 closes with the conclusions.

\section{Theory of detonations}
\label{sec:init}

\subsection{Detonation primer}
\label{ss:detonations}
In spite of the easy availability of good introductory and review material on detonations \citep[][i.e.]{fickett79,clavin04}, for the benefit of the reader we shall give a brief overview here.
A detonation in its simplest form is a shock that advances supersonically into a reactive medium, behind which chemical or nuclear reactions proceed. 
At least part of the region where exothermic reactions take place must be in sonic contact with the shock to inject energy and prevent the detonation from failing through dissipation. 
The Chapman-Jouguet (CJ) model for detonations \citep{chapman1899,jouguet05} is a one-dimensional model that describes the detonation in the limit of infinite reaction rate. In the state immediately behind the shock, the reactions are assumed to have progressed to completion. The Rankine-Hugoniot jump conditions, which express the conservation of mass, momentum, and energy flux across the shock front are simply modified by including the compositional change and the associated energy source term across the shock in the conservation equations. Assuming the upstream material to be at rest, the conservation equations accross the detonation front are:
\begin{eqnarray}
\rho_1 D &=& \rho_2 (D-u_2) \label{eq:dens} \\
P_1 + \rho_1 D^2 &=& P_2 +  \rho_2 (D-u_2)^2 \label{eq:mom} \\
(\varepsilon_1 + dq) + \frac{P_1}{\rho_1}+ \frac{1}{2} D^2 &=& \varepsilon_2 + \frac{P_2}{\rho_2}+ \frac{1}{2} (D-u_2)^2 \label{eq:ener}
\end{eqnarray}
where D is the detonation speed, $dq$ is the specific energy liberated in the burn, and the other quantities take on their usual meanings. Using the specific volume $V_i = 1/\rho_i$, and eliminating the fluid velocity of the ash state, $u_2$, from equations~\ref{eq:dens} and \ref{eq:mom}, we get an expression for the so called Rayleigh\footnote[1]{Also known as the Mikhel'son (sometimes spelled Michelson) line after the Russian physicist.} line:
\begin{equation}
\label{eq:rayleigh}
D^2 = \frac{P_2-P_1}{V_1-V_2} V_1^2
\end{equation}
Using eq.~\ref{eq:rayleigh} and either \ref{eq:dens} or \ref{eq:mom} one can easily eliminate $D^2$ and $(D-u_2)^2$ from the energy conservation equation to get an expression for the Hugoniot curve:
\begin{equation}
\label{eq:hugoniot}
2 (\varepsilon_1 + dq - \varepsilon_2) = (P_2+P_1)(V_2-V_1)
\end{equation}
\citet{zeldovich40},\citet{vonNeumann42} and \citet{doering43} improved the one dimensional description of detonations in what has become known as the ZND structure. They model a detonation as a one dimensional leading shock wave moving at the detonation speed trailed by a reaction zone of finite width in which energy is released and the fuel transforms into the burning products. The flow is assumed steady and planar. Solving the Rankine-Hugoniot equations without allowing for compositional change across the shock gives the thermodynamic conditions immediately behind the shock, the so called von Neumann state. From the von Neumann state on downstream the spatial variations of the hydrodynamic and thermodynamic variables in the reaction zone are determined by differentially following the compositional transmutations and the associated heat release. The ZND model has the advantage that it gives a prescription for calculating the width of the detonation wave (i.e. the distance between the shock and the final state of the ash). 
For one particular detonation speed $D_{CJ}$, the Rayleigh line is tangent (i.e. one point of intersection) to the Hugoniot curve in the $P-V$ plane, and the detonation is of so called CJ-type. In this case, the reactions terminate at the sonic point, that is, $D-u_2 = c_s$, where $c_s$ is the sound speed in the ash. If the final ash composition is known, this uniquely determines the propagation speed. It turns out that this so called CJ-speed corresponds to the smallest possible detonation speed, and many detonations fall into this category (but see also section \ref{ss:geo_curv}). For a detonation speed $D > D_{CJ}$, the Rayleigh line intersects the Hugoniot curve in two distinct places. The two points of intersection correspond to the ``weak'' or ``under-compressed'' and the ``strong'' solution. 
Strong detonations, which display the higher pressure increase and compression of the two solutions, have the reaction products moving subsonically with respect to the shock. Such detonations are often also referred to as ``overdriven", with overdrive factor $f = \big(\frac{D}{D_{CJ}}\big)^2$ \citep[e.g.][]{hwang00}. 

It is known that no steadily propagating weak detonation is possible \citep[e.g.][]{fickett79}. Supersonic and shockless weak detonations however do occur as a transient during the early stages of the initiation of detonations from shallow induction time gradients where they play an integral part in the formation of the ultimately steadily propagating strong or CJ detonation structure \citep{kapila89,short02,kapila02}.

While many of the observed properties of detonations can be explained by the simple one dimensional CJ and the slightly more advanced ZND models of steady state detonations,it is clear today that real detonations are very complex multi-dimensional structures that are at most steady state in an average sense. 
The structure and instabilities of fully developed detonations are very complex and have received considerable attention from researchers in both the astrophysical community
\citep[e.g.][]{gamezo99,sharpe99,timmes00a} and especially the combustion community \citep[e.g.][]{erpenbeck64,erpenbeck66,erpenbeck70,clavin97,short98,short03}. 
\subsection{Initiation of detonations}
\label{ss:det_init}
The question of how to form or initiate a detonation is similarly still an active area of both experimental, theoretical, and numerical research. 
The different ways of detonation formation can be grouped into two distinct categories: 
\begin{enumerate}
\item Direct initiation 
\item Spontaneous (initially shock-less) initiation.
\end{enumerate}
Direct initiation involves a blast wave or an otherwise formed shock propagating into a reactive medium at a speed exceeding the CJ speed for the conditions in the fuel. The resulting overdriven shock reaction-zone complex transitions or relaxes under the right conditions into a self supporting detonation \citep[e.g.][]{body97}. 

Spontaneous initiation, which this work focuses on, involves shock formation due to a spatial gradient in the initial conditions of the induction times of the fuel. It should be noted that spontaneous initiation may still require an external shock to ``precondition" the fuel such that auto-ignition may proceed from then on. The difference to direct initiation is that it is not the original shock that transitions into the leading shock of the detonation, but rather a new shock that forms when the nuclear fuel runs away coherently. The gradient mechanism of the initiation of a detonation was first proposed by \citet{zeldovich70}. The principal feature of this mechanism is the presence of a gradient in induction times (self-ignition delay times) that leads to a supersonic reaction wave. In the spontaneous wave picture, the phase velocity of the burning front is given by:
\begin{equation}
v_{ph} = \big(\frac{d\tau_{i}}{dr}\big)^{-1}
\end{equation}
 This wave transitions to a detonation when its phase velocity becomes equal to the velocity of a Chapman-Jouguet (CJ) detonation in that material. Zel'dovich's spontaneous wave concept, which ignores non-linear gas dynamical evolution and derives the burning wave speed from the initial conditions of the initiating center, describes the initiation of a detonation very well for shallow gradients. For steep gradients near criticality, however, the nonlinear gasdynamical effects are large and essential to the problem \citep{kapila02}. 

 \citet{lee78} build on Zel'dovich's idea and publicized the generalized picture of shock wave amplification through coherent energy release (SWACER). \citet{bartenev00} summarize the key features of the mechanism as follows:

\begin{itemize}
\item Gas layer(s) with minimal induction time(s) ignite(s) first
\item Shock propagates to adjacent layer(s), which are on threshold of ignition.
\item Shock initiates instantaneous explosion of layer, which strengthens shock.
\end{itemize}

In the astrophysical context of thermonuclear supernovae, \citet{blinnikov86} and \citet{blinnikov87} were among the first to discuss the initiation process of detonations from a temperature gradient in degenerate \nuc{12}{C}/\nuc{16}{O} matter. 
\citet{khokhlov97} studied induction time gradient initiations for detonations that were not solely due to temperature but also fuel concentration.
Pioneering work for determination of critical radii in the context of SNe Ia  has been performed by \citet{arnett94b}, followed by \citet{niemeyer97} and \citet{roepke07}. 
This paper extends their work to include a study of the dependences of the critical gradients on composition, background temperature, peak temperature, geometry, and functional form of the temperature disturbance. Next, we discuss each of these briefly to motivate the components of our study.

\subsection{Fuel density}
The higher the fuel density, the smaller are the characteristic length scales for detonation. This is essentially due to the fact that thermonuclear reaction rates scale as the product of the number densities of the reactants (i.e. $r_{i,j}=n_i n_j <\sigma v>_{i,j}$). As a result of the higher reaction rates, burning time scales are shorter and the width of a fully developed self-supported \nuc{12}{C}/\nuc{16}{O} detonation wave therefore decreases with increasing density \citep[e.g.][]{khokhlov89}. The critical length scales (e.g. critical radii) for the initiation process are correspondingly smaller as well. At low densities, the critical radius becomes larger than the density scale height of the star, leaving constant density simulations rather meaningless. Nevertheless, the density (for a given choice of peak and background temperature) at which the critical radius becomes equal to the size of the white dwarf can be considered a very conservative estimate of the lowest density at which a gradient initiated detonation could be successfully launched. A fiducial density of $10^7 \gcc$ is adopted here for most cases, but densities as low as $10^6 \gcc$ and as high as $3 \times10^7 \gcc$ are considered.

\subsection{Temperature profiles}
Past work in the determination of critical sizes of gradient initiated detonations in WD environments all considered linear temperature gradients and a cold background. 
Typically, for a given spatial extent of the temperature disturbance, the smallest peak temperature that would still lead to detonation in a cold surrounding medium was determined. 
A slightly different approach is taken here: For a given peak temperature the smallest size of the heated region that still leads to detonation is determined (see section \ref{sec:critital_radii}).

The choice for the linear profile of the temperature gradient seems arbitrary. For unresolved simulations, a hypothetical Gaussian or exponential temperature gradient on the sub-grid scale is at least as plausible as a linear one. Whether or not there is considerable dependence of the detonation conditions on the choice of functional form for the temperature gradient is therefore a question well worth investigating. Since the initiation site is possibly embedded in an environment where even the asymptotic background temperature has been heated to an appreciable temperature (by a shock for example, or by compression), a range of ambient background temperatures is examined here.

\begin{figure}
\plotone{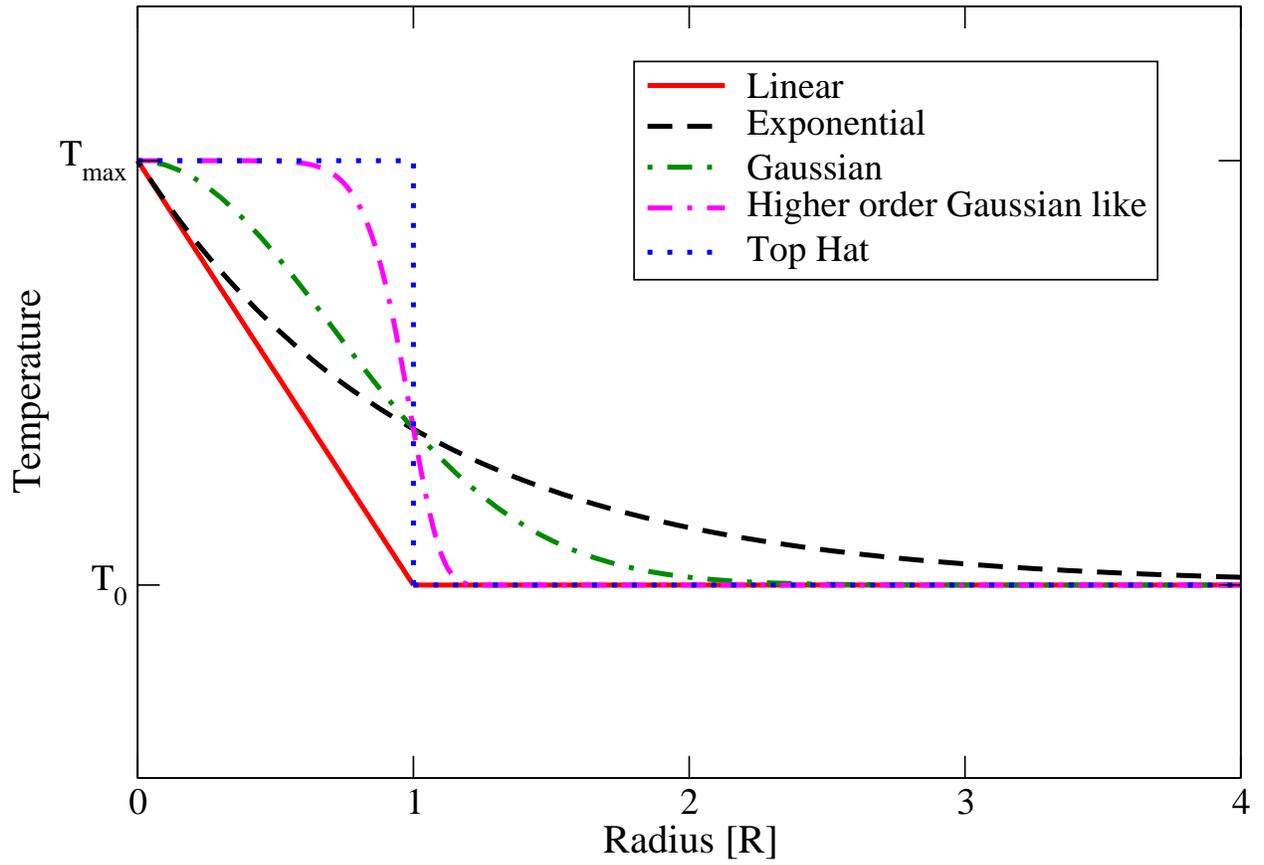}
\caption{\label{fig:tprof} Initial temperature profiles considered in this work.}
\end{figure}

The initial temperature profiles of the simulations consist of a hotspot with peak temperature $T_{max}$ falling off to an ambient temperature $T_0$ \mbox{(see fig.~\ref{fig:tprof}).} The functional form of the temperature perturbation is either linear

\begin{equation}
T(r)= 
\left\{
\begin{array}{ll}
T_{max} - \frac{T_{max}-T_0}{R} r & \textrm{for $ r \le R$,}\\
\\
T_0 &\textrm{for $r > R$,}
\end{array} \right.
\end{equation} \\
exponential
\begin{equation}
T(r) = (T_{max} - T_0)  \exp(-r/R) + T_0,
\end{equation} \\
Gaussian
\begin{equation}
T(r) = (T_{max} - T_0)  \exp[-(r/R)^2] + T_0,
\end{equation} \\
higher order Gaussian like (hereafter referred to as g10)
\begin{equation}
T(r) = (T_{max} - T_0)  \exp[-(r/R)^{10}] + T_0,
\end{equation} \\
or a top hat 
\begin{equation}
T(r)= 
\left\{
\begin{array}{ll}
T_{max} & \textrm{for $ r \le R$,}\\
\\
T_0 &\textrm{for $r > R$.}
\end{array} \right.
\end{equation} \\
Depending on whether the problem is setup in planar or spherical geometry, $r$ is the distance from a reflecting boundary or the origin respectively.

\subsection{Composition}
SNe Ia progenitors are likely \nuc{12}{C}/\nuc{16}{O} white dwarfs. The \nuc{12}{C} to \nuc{16}{O} ratio in a massive white dwarf at the end of core \nuc{4}{He} burning is not very well constrained. The large uncertainty of one of the most important nuclear reaction rates,  $\nuc{12}{C}(\alpha,\gamma)\nuc{16}{O}$, translates into a range of possible \nuc{12}{C} to \nuc{16}{O} ratios. 
 In this work, we explore the sensitivity of the critical sizes on the \nuc{12}{C} to \nuc{16}{O} ratio for a range of allowed ratios (see fig. 4 from \citet{dominguez01}).

The details of the accretion process towards a Chandrasekhar mass white dwarf in the single degenerate channel are not solved, but the possibility of a significant layer or at least admixture of \nuc{4}{He} in the outer layers of the massive white dwarf remains. We would like to stress that the rate of the \mbox{$\nuc{12}{C} (\alpha,\gamma) \nuc{16}{O} $} reaction proceeds much faster than carbon burning. Therefore, adding \nuc{4}{He} into the mixture will lead to a more reactive medium and significantly decrease the critical radii. In this work, we explore the sensitivity of the critical sizes on the possible admixture of \nuc{4}{He}.

\subsection{Geometry and curvature}
\label{ss:geo_curv}
Planar detonations in \nuc{12}{C}/\nuc{16}{O} matter are known to be of pathological type for densities $\rho > 2 \times 10^7$ g cm$^{-3}$ \citep{khokhlov89,sharpe99,gamezo99}.  Endothermic photo-disintegration reactions give rise to a frozen sonic point in the reaction zone of the detonation at which the local sound speed equals the velocity of the flow in a frame in which the leading shock of the detonation is stationary. Material downstream of the sonic point is moving supersonically away form the shock front and is out of causal contact with the leading edge of the detonation. The sonic point, also known as the pathological point, lies for CO WDs generally in the oxygen or silicon burning layer. The detonation speed of a pathological detonation is an eigenvalue of the steady equations \citep[e.g.][]{yao06} and it is larger than the detonation speed of the corresponding CJ detonation. 
 
Steady planar detonations in \nuc{12}{C}/\nuc{16}{O} at lower density are of the CJ-type, frozen subsonic throughout all the burning zones with the sonic point located at the end of the reaction zone where nuclear statistical equilibrium is reached. It is well known, however, that even a small amount of curvature can significantly influence the propagation speed and structure of detonations in SNe Ia environments. The curvature causes \nuc{12}{C}/\nuc{16}{O} detonations to be of pathological type even for densities as low as $\rho = 5 \times 10^6$ g cm$^{-3}$. For curvatures larger than $1.5 \times 10^{-11}$ cm$^{-1}$, the pathological sonic point falls near the end of oxygen burning, greatly affecting the length scales of carbon, oxygen, and silicon burning stages of the detonation \citep{sharpe01}.  The curvature constraint means in practice that all detonations that initiate from a localized hot spot near the surface of a white dwarf for densities $\rho\sim10^7 \gcc$ are of pathological type.  

\citet{he94} show for direct initiation, that non-linear curvature effects cause the critical radius R$_c$, in this case defined as the radius corresponding to the smallest possible igniter energy at which the shock wave of the Sedov problem  \citep{sedov59} transitions or relaxes to a wave propagating at the CJ-velocity of the mixture, is about 10 times larger for spherical geometry when compared to planar. It is intuitively clear that the critical radii for spontanous initiation in 1D spherical geometry will be larger compared to 1D planar geometry. To quantify the differences critical radii in both geometries are determined for the same composition, density, background temperature and peak temperature.   

\subsection{Multi-Dimensional Effects}
Real detonations are not one dimensional objects. They rather exhibit a multi-dimensional cellular structure with complicated internal substructures that includes transverse shocks, triple points (points of maximum pressure where transverse shocks intersect the Mach stems) and weak incident shocks \citep{timmes00a,fickett79}. The cellular structure behind the detonation front results in pockets of incompletely burned fuel which leads to a reduction in energy input compared to the one dimensional case and results in a reduced detonation velocity \citep{boisseau96}. 

While the cellular structure is important for the width of a self sustaining steady detonation and the resulting pockets of inhomogeneously burned fuel will likely leave a signature in the spectra of SNe Ia \citep{gamezo99}, it is unclear how it affects the critical radii for initiation of detonations via the gradient mechanism. 
An attempt to address this question was made by determining critical radii for a few select cases in 2D. Significant differences in the critical radii when going from 1D to 2D were not noted. The results however, should be considered inconclusive, since the simulations started from idealized perfectly smooth initial conditions. Without initial perturbations, the time scale for the cellular instabilities to develop from numerical round off errors is large compared to the time scale for oxygen ignition. It is conceivable that slightly more realistic noisy initial conditions will facilitate faster growth of the cellular instability and change the picture altogether. Furthermore, it is conceivable that one has to spatially resolve the carbon burning layer to see a substantial manifestation of higher dimensional effects. This question certainly deserves further detailed study, which unfortunately goes beyond the scope of this article.

\section{Critical Size Determination}
\label{sec:critital_radii}
Past efforts in determining minimal critical length scales for the initiation of detonations in white dwarf environments include \citet{arnett94b}, \citet{niemeyer97}, and \citet{roepke07}. All of these works are one dimensional simulations.
\citet{arnett94b} determined critical radii in \nuc{12}{C}/\nuc{16}{O} material by determining the smallest spheres with a peak temperature \mbox{$T_{max} = 2.0 \times 10^9$ K} and a background temperature \mbox{$T_0 =2.0 \times 10^8$ K} and a top hat temperature profile that would lead to a runaway. A successful detonation was declared if the heated material was completely burned within 10 sound crossing times (1000 time steps). 
Although a thermonuclear runaway of the heated region is a necessary ingredient for the formation of a detonation, it is by no means sufficient. Additionally to the runaway, a  shock-reaction zone complex needs to form and survive. The functional form of the temperature profile and the extent of the heated region both play a significant role here. 

To illustrate, a hot spot in 1D planar geometry with a top hat temperature profile for $\rho = 1.0 \times 10^7 \gcc$, \mbox{$T_{max} = 2.8 \times 10^9$ K}  and \mbox{$T_0 = 1.0 \times 10^7$ K}, does not lead to a detonation, no matter how large the heated region is. While a runaway occurs and quickly consumes all the carbon and oxygen, the resulting shock wave is too weak to form a detonation (see also table 1 and the discussion in \citet{khokhlov91a}). In fact, owing to the lack of gradients in induction time, the top hat profile turns into a problem of direct initiation. The heated region runs away isochorically, turning the initial conditions into a Sedov problem. The strength of the generated shock wave is independent of the size of the heated region, leading to the result that for a top-hat temperature profile, a larger heated region (once above a minimum critical size) will not make a detonation more probable. In fact, for a while it was argued that for this reason the initiation of detonations in degenerate \nuc{12}{C}/\nuc{16}{O} mixtures is impossible \citep[e.g.][]{nomoto76, mazurek77}. 
This claim, however, was based on the assumption of direct initiation. It was shown later that formation of detonations in degenerate \nuc{12}{C}/\nuc{16}{O} matter was still possible via the spontanous mode of initiation, e.g. from a suitable temperature gradient \citep{blinnikov86,blinnikov87}. 

 In contrast, for the same parameters a linear temperature profile leads to detonation at $R_{crit} = 1.2$ km (see table~\ref{tab:lin_plan}). 
This example explicitly demontrates that there is no such thing as a ``critical mass'' of fuel that, if heated above a certain temperature, would necessarily result in a successful detonation. 
Instead, the details of the temperature profile are essential for the outcome of the initiation problem. 
Therefore, the radii and masses of \mbox{table 1} in \citet{arnett94b} are to be interpreted as lower bounds, constraining the conditions for a successful thermonuclear runaway - a necessary, but not sufficient, condition for the initiation of a detonation. 
 Indeed, the critical radii for \mbox{$\rho = 1.0 , 3.0 \times 10^7 \gcc$} and  \mbox{$T_{max} = 2.0 \times 10^9$ K} obtained here are about an order of magnitude larger. (The value  \mbox{$T_{0} = 2.0 \times 10^8$ K} used in their work is bracketed by two of the choices for the background temperature used in this work \mbox{$T_{0} = 1.0 \times 10^7$ K} and \mbox{$T_{0} = 4.0 \times 10^8$ K.}) 

\subsection{Problem setup and method}
\label{subsec:setup}
This work uses the \FLASH\ code \citep{fryxell00} for the numerical experiments. 
\FLASH\ is an Eulerian code with adaptive mesh refinement (AMR) capabilities that solves the compressible reactive Euler equations with a directionally split implementation of the piecewise parabolic method (PPM) \citep{colella84}. 
The Riemann solver is implemented in a way to handle a general non-polytropic equation of state (EOS) \citep{colella85}. 
The EOS is appropriate for the compositions, densities, and temperatures encountered here \citep{timmes99a,timmes00b,fryxell00}. 
For the nuclear energy release we use an inexpensive 13 species $\alpha$-chain plus heavy ion nuclear reaction network containing \nuc{4}{He}, \nuc{12}{C}, \nuc{16}{O}, \nuc{20}{Ne}, \nuc{24}{Mg}, \nuc{28}{Si}, \nuc{32}{S}, \nuc{36}{Ar}, \nuc{40}{Ca}, \nuc{44}{Ti}, \nuc{48}{Cr}, \nuc{52}{Fe}, and \nuc{56}{Ni} \citep{timmes99b,timmes00c}. 
As suggested in \citet{fryxell89}, nuclear burning was suppressed in shocks, since the real shock is much thinner in spatial extent compared to the one spread out over 2-3 zones by PPM. 
The shock was defined as a region of compression (negative velocity divergence) and a significant pressure jump ($\Delta P/P > 1/3$). 

The time step was chosen to be $min(t_{hydro},t_{burn}) = min(0.8 \; CFL,0.01 \;  E_{int}/\epsilon_{nuc})$, where $E_{int}$ is the specific internal energy [erg g$^{-1}$], $\epsilon_{nuc}$ is the specific nuclear energy generation rate  [erg g$^{-1}$ s$^{-1}$] and $CFL$ is the Courant-Friedrichs-Lewy number \citep{courant28}. Limiting the energy added to a computational cell in one time step to 1\% of the internal energy of that cell is supposed to improve the feedback of the nuclear burning on the hydrodynamics, which are handled in an operator split way. 
The factor 0.01 is chosen somewhat arbitrarily; a smaller factor would be more conservative, but also make the simulations computationally more expensive. A choice of time step purely based on the $CFL$ number would ignore important coupling of the nuclear burning to the hydrodynamics and facilitate the spurious initiation of detonations. 

The initial conditions consisted of constant density and composition throughout the computational domain.
The spatial extent of the computational domain was chosen to be $64 \times R$, which were empirically found to be large enough to cover the locus of oxygen ignition for all densities, compositions, and geometries considered. 64 top level blocks with 16 computational zones each and 7 levels of refinement were consistently used. This particular setup means that $R$ was resolved equally in all the runs, whereas the spatial resolution ($R/1024$) varied in absolute terms from case to case.
The simulation time was dependent on the particular density, but most simulations were run out to ~50,000-100,000 time steps.   

Once a radius $R^{+}$, for which a detonation ensued, and a radius $R^{-}$, for which failure was the result, were determined, the critical radius was found by bisection. A trial radius $R_{try} = \frac{R^{+} + R^{-}}{2}$ was chosen, and if it led to success (failure) then it replaced the old $R^{+}$ ($R^{-}$).  $R^{+}/R^{-} < 1.1$ was the minimal goal, but often the radii were determined to 2 significant figures. This means that each determination of a critical radius involves $\sim10$ runs. The smallest (largest) $R^{+}$ ($R^{-}$) is listed as $R_{c,max}$ ($R_{c,min})$ in the tables. 

A code to code comparison with \citet{niemeyer97}, who used the Lagrangian hydrodynamics code KEPLER with a 19 nuclide network and comparable zoning, was performed for their three published low density cases ($\rho \leq 10^8 \gcc$) that  have a 50/50 \nuc{12}{C}/\nuc{16}{O} composition. 
Just as in their work, $T_{max} = 3.2 \times 10^9$ K, $T_{0} = 4.0 \times 10^8$ K and a linear temperature gradient were selected.
In all three cases, the critical radii determined here agree exquisitely with those found by \citet{niemeyer97} (see table~\ref{tab:comparison}), whose Lagrangian calculations were also unresolved.

\subsection{Criteria for success or failure}
\begin{figure}
\plotone{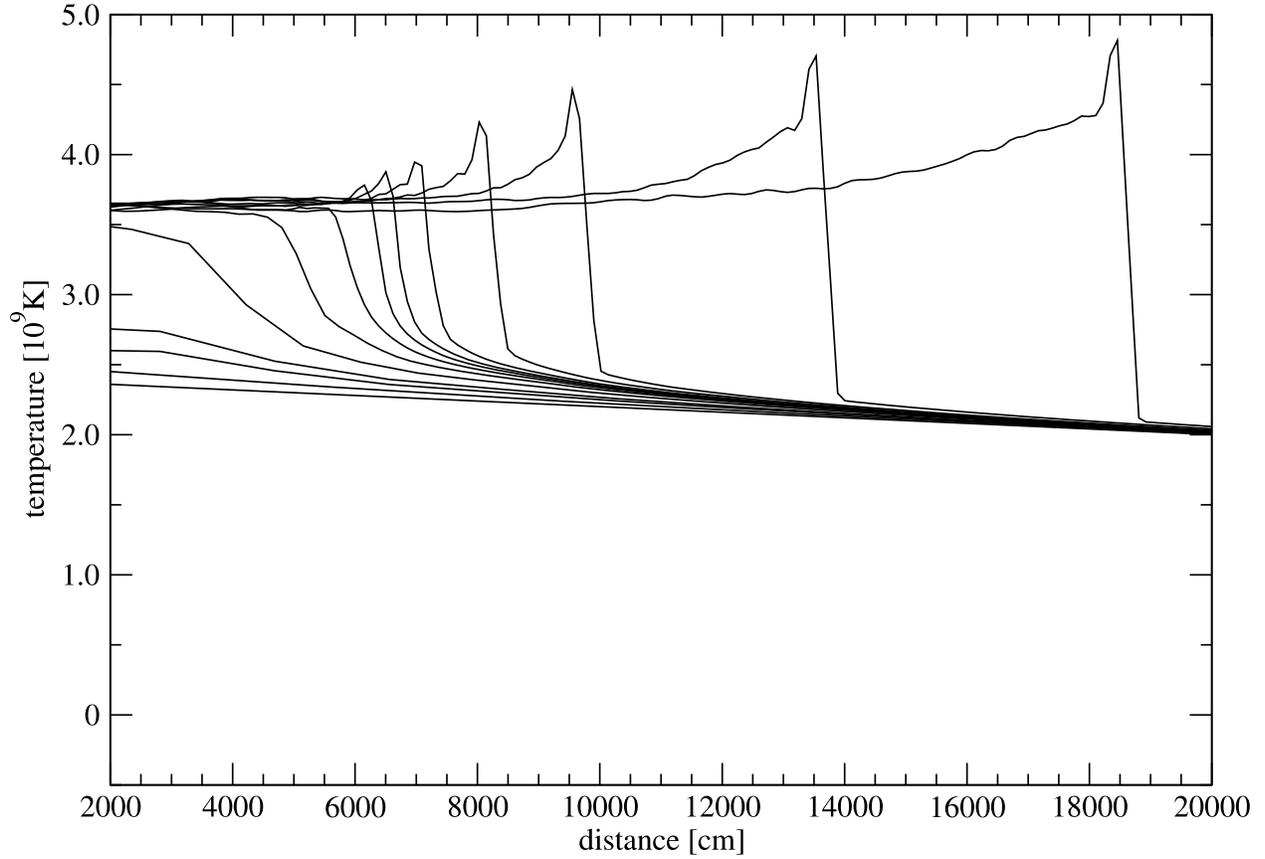}
\caption{\label{fig:1point2} Shock formation during the early stages of the runaway. The density is $10^7$ g cm$^{-3}$ in planar geometry for initial conditions that successfully initiate a detonation. The linear temperature disturbance has $R =  1.2$ km, \mbox{$T_{max} = 2.4 \times 10^9$ K} and \mbox{$T_{0} = 1.0 \times 10^7$ K.} See also figs. \ref{fig:early_xc12}, \ref{fig:early_pres} and \ref{fig:8panels}.}
\end{figure}

\begin{figure}
\plotone{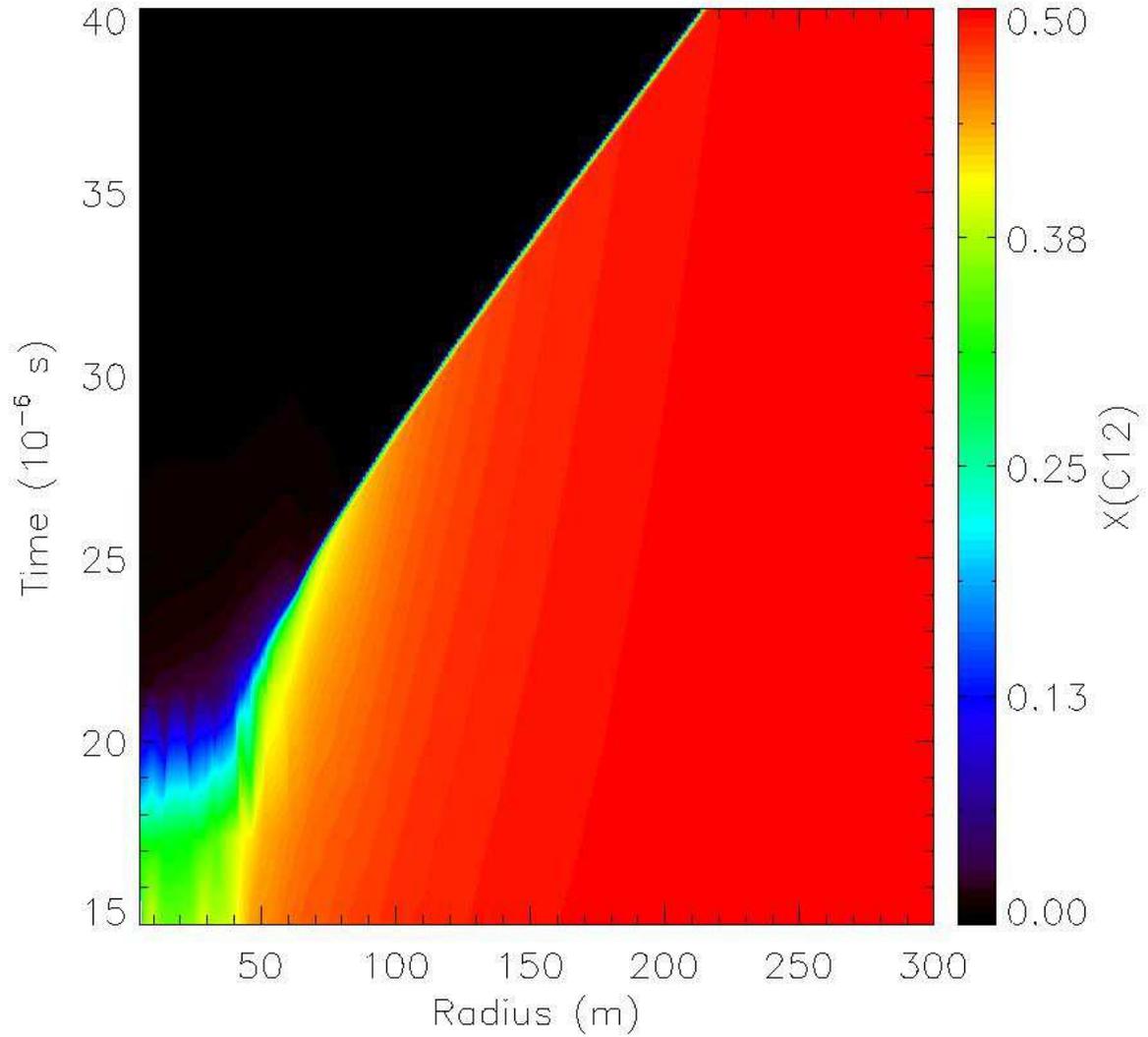}
\caption{\label{fig:early_xc12} \nuc{12}{C} mass fraction in a space time diagram during the early stages of shock formation. Clearly visible is the initial almost synchronous burning of a narrow boundary layer, followed by a smeared out reaction front which accelerates to form the detonation. The initial conditions are as in fig. \ref{fig:1point2}.}
\end{figure}

\begin{figure}
\plotone{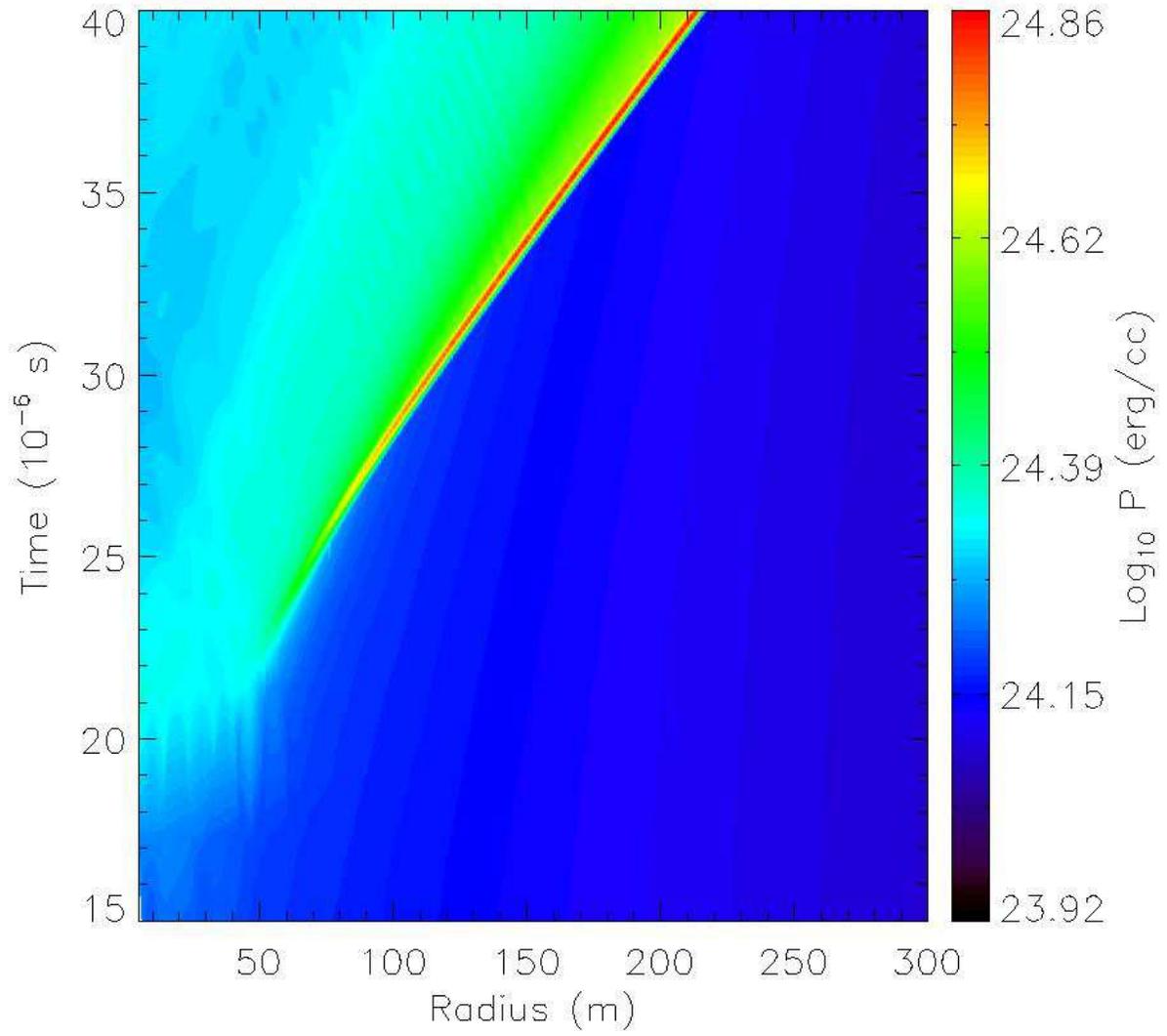}
\caption{\label{fig:early_pres} Pressure during the early stages of shock formation. Visible is the increase in propagation velocity of the location of the pressure jump between $\sim 60$ m and $\sim 120$ m. The initial conditions are as in fig. \ref{fig:1point2}.}
\end{figure}

\begin{figure}
\epsscale{0.8}
\plotone{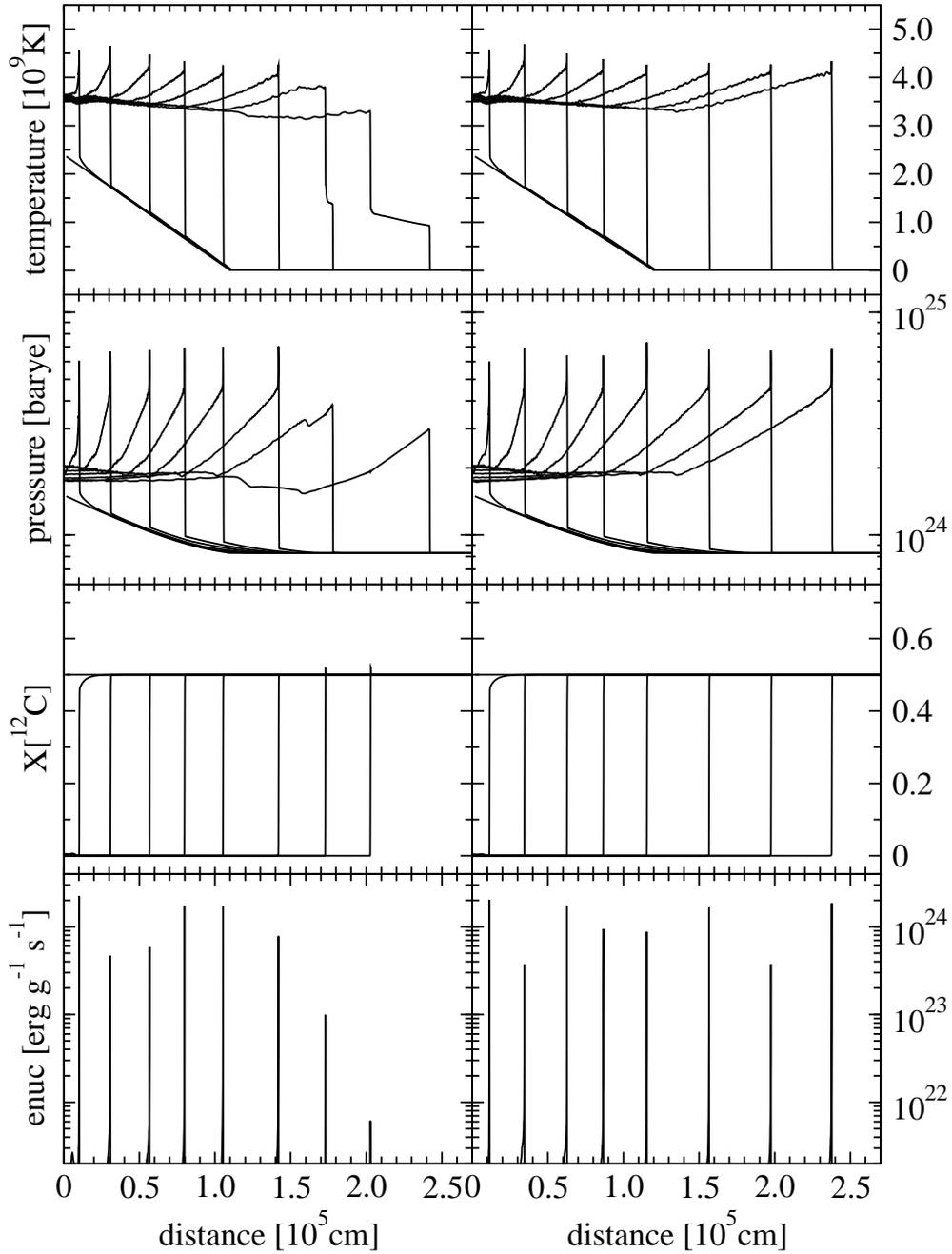}
\caption{\label{fig:8panels} Shown are time series of snapshots of temperature, pressure, mass fraction of \nuc{12}{C}, and nuclear energy generation rate at a density of $10^7$ g cm$^{-3}$ in planar geometry for initial conditions that fail to lead to a detonation ($R = 1.1$ km, left) and initial conditions that successfully initiate a detonation ($R =  1.2$ km, right). The linear temperature disturbance has \mbox{$T_{max} = 2.4 \times 10^9$ K} and \mbox{$T_0 = 1.0 \times 10^7$ K.} }
\end{figure}

The method described above for determining the critical radii relies on a criterion for success or failure. The decision was made by inspecting the thermodynamic, hydrodynamic and compositional profiles of the initiation simulations at late times. 
For radii near criticality, the evolution of the sub- and supercritical cases initially proceeds in a very similar way. The temperature rather uniformly rises in the innermost (originally most pre-heated) region where \nuc{12}{C} is depleted first (see figs.~\ref{fig:1point2} and~\ref{fig:early_xc12}). The rapid almost isochoric burning in the thin boundary layer surrounding the peak of the temperature inhomogeneity leads the formation of a shock (see figs.~\ref{fig:early_pres} and~\ref{fig:1point2}), which further steepens via the SWACER mechanism (see section~\ref{ss:det_init}). The shock then propagates down the temperature gradient, immediately trailed by a narrow region of carbon burning (see fig.~\ref{fig:8panels}), accelerating to detonation speed.

\begin{figure}
\epsscale{1.0}
\plotone{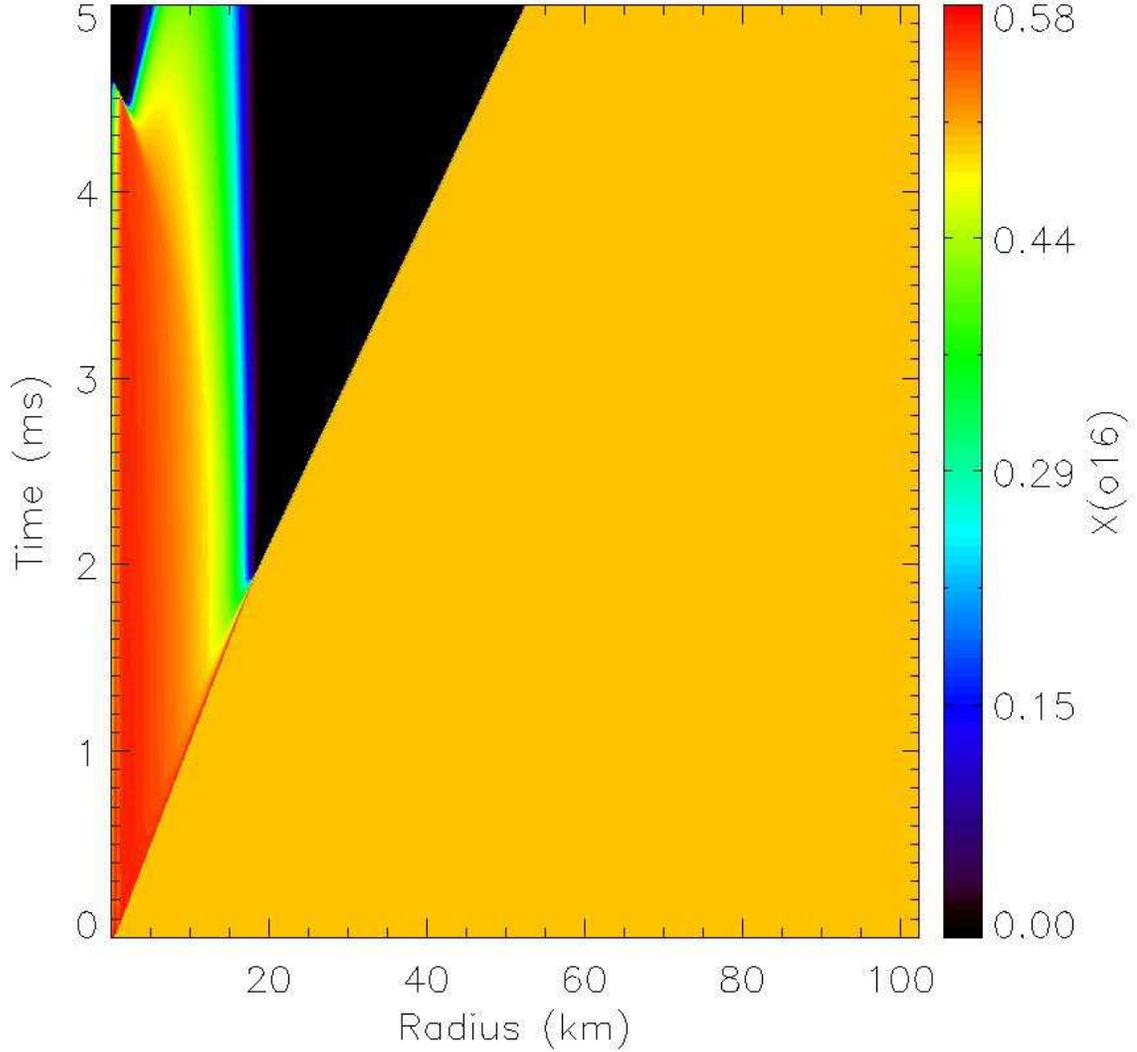}
\caption{\label{fig:o16} Space-time diagram of the mass fraction of \nuc{16}{O} for a succesful detonation in spherical geometry for an initial composition of equal parts \nuc{12}{C} and \nuc{16}{O} by mass. The radius of the initial linear hot-spot is $R=1.6$ km and \mbox{$T_{max} = 3.2 \times 10^9$ K,} \mbox{$T_0 = 4.0 \times 10^8$ K,} \mbox{$\rho = 10^7 \gcc$.} The increase in detonation speed is clearly visible as oxygen ignites after \mbox{$\sim~2$ ms.}  }
\end{figure}

\begin{figure}
\plotone{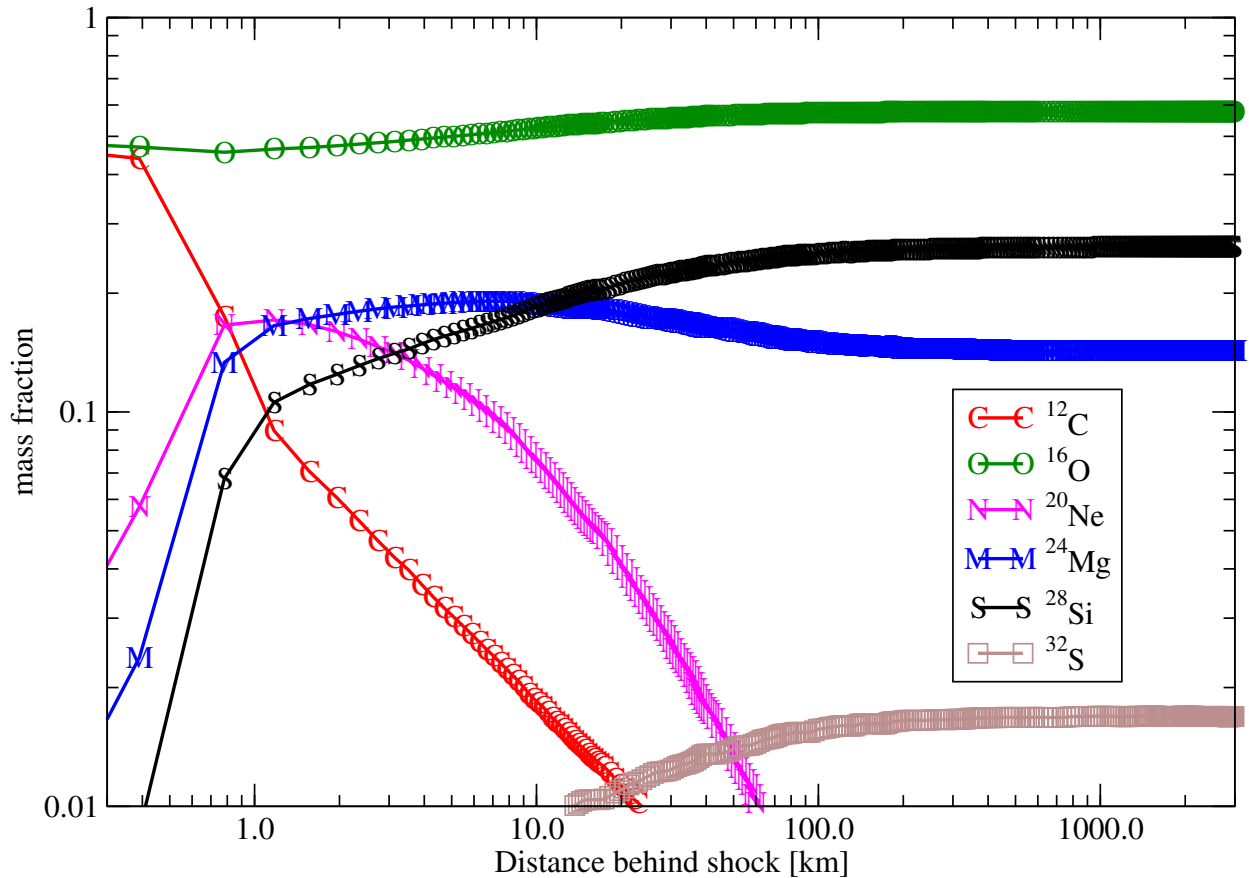}
\caption{\label{fig:cdet} Shown are mass fraction of select nuclear species as a function of distance behind the leading shock.  The fuel density is $\rho=10^6 \gcc$. The detonation successfully propagates consuming the carbon fuel, producing intermediate mass elements (more oxygen, magnesium, silicon and sulfur). Their mass fraction appear to reach asymptotic levels far behind \mbox{($\sim100$ km)} the leading edge of the detonation. Similar results were obtained by \citet{imshennik84} and \citet{khokhlov89b}.}
\end{figure}

While the early evolution  of the barely sub- and supercritical cases is nearly identical, the carbon burning region fails to couple to the leading shock and falls behind (see fig.~\ref{fig:8panels} left). Once the advancement of the leading edge of the carbon burning zone had ground to a halt the detonation was declared a failure.  A success is relatively more difficult to declare, as the decoupling of the shock and the carbon burning region may occur well outside the originally pre-heated region (see fig.~\ref{fig:8panels} left). No ``failures" however were ever observed to occur after the successful ignition of oxygen, which due to the longer burning time scale is delayed compared to carbon burning (see the space time diagram fig.~\ref{fig:o16}). Oxygen ignition was therefore taken as the criterion for success. At the lowest densities ($\rho = 10^6 \gcc$), the curious situation arises that the temperature reached after carbon burning falls below the oxygen ignition threshold. Detonations propagating steadily for distances exceeding the size of the star are observed that fuse the carbon and the carbon burning product neon, but fail to fuse oxygen (see fig.~\ref{fig:cdet}). If the carbon burning region remained coupled to the shock for for a distance that approached a considerable fraction of the size of the star ($\sim 2000$ km), then the initiation was also considered a ``success".

\subsection{Resolution study}
\label{ss:res_study}
\begin{figure}
\plotone{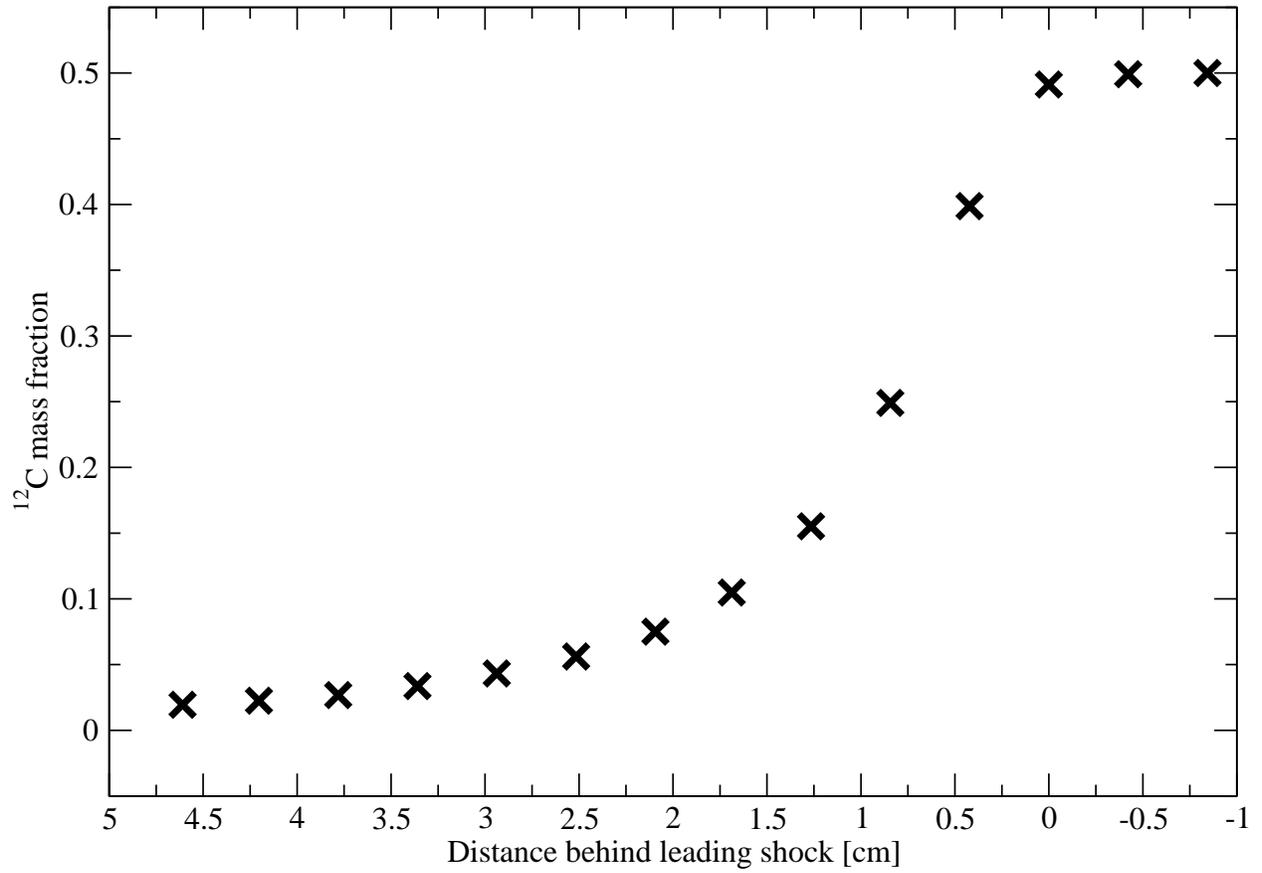}
\caption{\label{fig:sh_c12_lr15} Resolved \nuc{12}{C} burning zone trailing the leading shock in a fully developed detonation for $\rho=10^7 \gcc$ and $T_0=1.0\times10^9$ K. The data are from the run with 15 levels of refinement listed in table~\ref{tab:res_study}. There is one cross symbol for each computational cell.}
\end{figure}

Ideally, a resolution high enough to guarantee convergence would have been used. Extrapolating somewhat freely from the work by \citet{hwang00}, who perfomed a study on numerical resolution of detonation waves using a higher order essentially non-oscillatory hydrodynamics scheme, it was expected that convergence may be reached no sooner than at a resolution of  at least 40 zones per carbon burning length scale. A resolution that high was unfortunately prohibitively expensive for the large number of detonation initiation cases intended for study. The critical sizes presented in this paper were determined with simulations that left the region of carbon burning unresolved. To get at least some idea about the impact of unresolved scales on the critical sizes, a resolution study was performed. For the resolution study somewhat arbitrarily a 50/50 \nuc{12}{C}/\nuc{16}{O} composition with a linear temperature gradient at a density of  \mbox{$10^7 \gcc$}, a peak temperature \mbox{$T_{max} = 2.4 \times 10^9$ K} and a background temperature \mbox{$T_0 = 1.0 \times 10^9$ K} in planar geometry were chosen. For these conditions, the carbon burning length scale is a few cm (see the resolved calculation depicted in fig.~\ref{fig:sh_c12_lr15}). The critical radii increase with increasing resolution and appear to reach an asymptotic value once the carbon burning length scale is resolved (see table~\ref{tab:res_study}). The main thing to take away from this study is that the critical radii determined here with unresolved simulations are, while at the right order of magnitude, not very accurate. In fact, the critical sizes of a corresponding resolved simulation appear to be a factor of two or three higher.

\section{Results}
In this section, the results for the determination of the critical radii (decay constants) are presented. To highlight the trends with and dependences on density, temperature, composition, and geometry, they are separated and described in turn.  
 
 \subsection{Fuel density}
\begin{figure}
\plotone{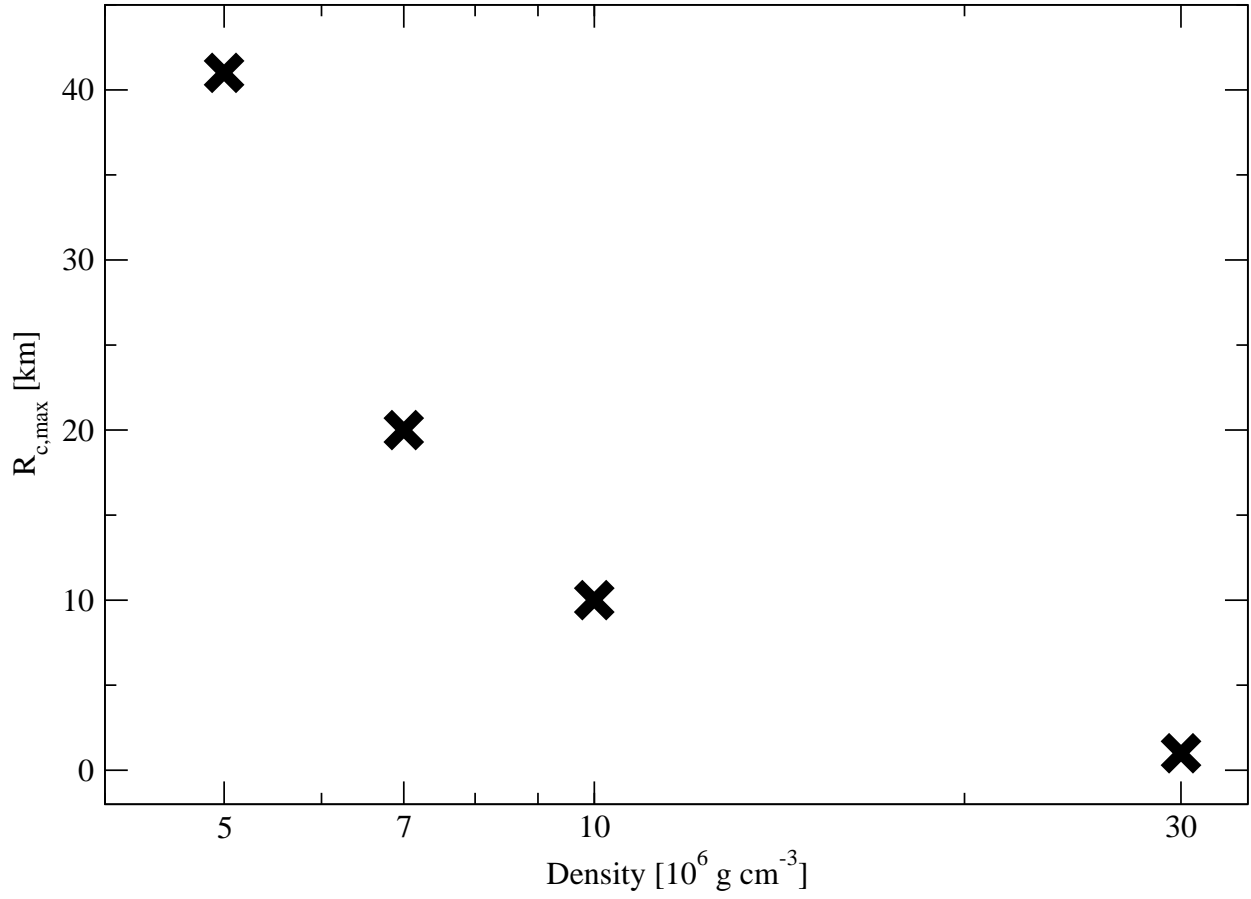}
\caption{\label{fig:dens} Variation of critical radii with density, all other variables held fixed. Chosen at random here $T_{max}=2.0\times10^9$ K, $T_{0}=1.0\times10^9$ K and a linear temperature profile in spherical geometry.}
\end{figure}

 As expected, the critical sizes for spontaneous initiation of detonations decrease with increasing density (see fig.~\ref{fig:dens}). The dependence on density is quite strong, with a density increase of less than an order of magnitude (going from $5.0\times10^6 \gcc$ to $3.0\times10^7 \gcc$), leading to a decrease in the critical radii around (and sometimes exceeding) two orders of magnitude (see table~\ref{tab:lin_sph}). 

Contrary to a statement in \citet{roepke07}, we find that in \nuc{12}{C}/\nuc{16}{O} material detonations can form at densities as low as \mbox{$\rho = 10^6$ g cm$^{-3}$}. However, to get a detonation at such low density from gradients smaller than the size of the star, it is necessary to have suitably high background temperatures (see table~\ref{tab:lin_plan}). Unlike at the higher densities, oxygen ignition is not required for success. The resulting detonations appear to propagate steadily for distances exceeding several thousand km, curiously only consuming the carbon without burning the oxygen (see fig.~\ref{fig:cdet}). An initially overdriven detonation that also burns the oxygen initially eventually relaxes into a steady propagating detonation that only burns carbon at this low of a density. 

\subsection{Temperature profile}
\subsubsection{Peak and background temperature}
Higher peak temperatures decrease the critical sizes for detonation. However, there is a limit to this trend. For a given $T_0$ and $\rho$, there is a $T_{max}$ above which the critical sizes appear to asymptote (see tables~\ref{tab:lin_sph} and \ref{tab:lin_plan}). 

Four background temperatures were chosen for the parameter study. They are $T_0=1.0\times10^7$ K (very cold), 
$T_0=4.0\times10^8$ K (to compare with \citet{niemeyer97}, cold), $T_0=1.0\times10^9$ K (hot), 
and $T_0=1.5\times10^9$ K (very hot).
The effect of the variation of the background temperature on the critical sizes is two-fold. 
\begin{enumerate}
\item Geometrical. For the linear profile for example, for a given $T_{max}$ the same slope clearly corresponds to different radii.  
\item Raised internal energy. A higher background temperature means that the internal energy of the fuel is higher, and a weaker shock is needed to ignite it. 
\end{enumerate}
 Both effects go the same way; the larger $T_0$, the smaller the critical radius. 
For the lower peak temperatures, the dominating effect is geometrical, since the slope of the critical profile is independent of $T_0$ (see table~\ref{tab:lin_sph}).
For the higher peak temperatures, the critical slopes increase with increasing $T_0$.  
\subsubsection{Functional form}

\begin{figure}
\plotone{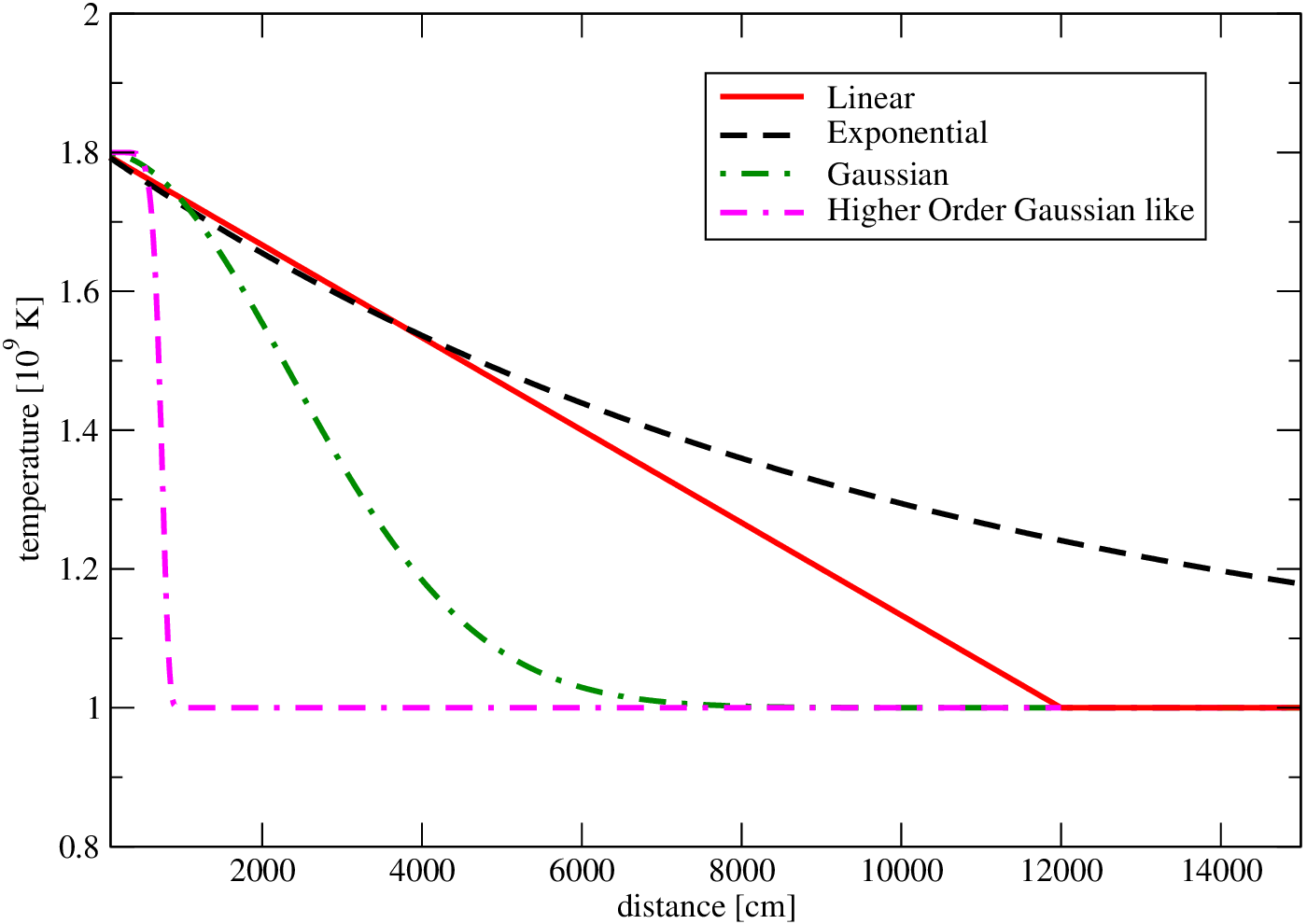}
\caption{\label{fig:func18} Critical  temperature profiles for \mbox{$T_{max} = 1.8 \times 10^9 $ K} and \mbox{$T_0 = 1.0 \times 10^9$ K.} The density is  \mbox{$10^7$ g cm$^{-3}$. } }
\end{figure}

The functional form of the temperature profile can have a very large influence on the outcome. The profiles can be separated into two categories:
\begin{enumerate}
\item Sharply peaked. Linear and exponential fall into this category. 
\item Flat topped. Gaussian and higher order Gaussian-like make up this category. 
\end{enumerate}
The critical exponential profile is such that the slope at the origin is very nearly equal (or slightly larger) to a critical linear profile for the same $T_{max}$ and $T_0$ (see figs.~\ref{fig:func18}, \ref{fig:func24} and table~\ref{tab:exp_plan}). This explicitly demonstrates that it is not the amount of mass heated above a certain temperature that is the deciding factor but rather the gradient in induction times. 

The critical flat topped profiles behave differently from the  critical sharply peaked profiles (see tables~\ref{tab:gauss_plan} and \ref{tab:g10_plan}). For small $T_{max}$, the flat topped profiles more readily lead to detonation (see fig.~\ref{fig:func18}). In fact, for a Gaussian profile, even peak temperatures as low as $T_{max}=1.6\times10^9$ K (for $\rho=10^7 \gcc$) initiate a detonation at relatively small spatial scales of $\sim10$ km (see  table~\ref{tab:gauss_plan}), which is far lower than the limit of $T_{max}=1.9\times10^9$ K given by \citet{roepke07}. For large $T_{max}$, the flat topped profiles require larger spatial scales to lead to detonation (see fig.~\ref{fig:func24}). At high temperatures, the flat topped profiles are disfavored due to their closer resemblance of a top-hat isochoric blast. The almost instantaneous runaway resulting from the high initial temperatures well above the threshold of rapid carbon fusion in the flat topped region leads to a synchronized burning of the carbon. The inability of such a constant volume explosion in CO matter to directly initiate a detonation is the reason why the flat topped profiles need to be ``stretched" out until the gradient mechanism can operate in the wing of the profile (see fig.~\ref{fig:func24}). The low peak temperature flat topped profiles do not share this problem since the initially small temperature gradient in the flat topped region is amplified during the longer lasting ``smoldering phase" leading up to the runaway. 

\subsection{Composition}
\subsubsection{\nuc{12}{C} / \nuc{12}{O} }
\label{subsec:co}
The mass fractions of \nuc{12}{C} and \nuc{16}{O} in putative SNe Ia progenitors is uncertain averaged over the star, and allows for even more variation near the surface. To ascertain the dependence of the  detonability on the \nuc{12}{C} to \nuc{16}{O} ratio, we have determined a few critical radii for carbon rich compositions consisting of 70\% \nuc{12}{C}, 30\%  \nuc{16}{O} and 60\% \nuc{12}{C}, 40\%  \nuc{16}{O} as  well as oxygen rich compositions consisting of 40\% \nuc{12}{C}, 60\%  \nuc{16}{O} and 30\% \nuc{12}{C}, 70\%  \nuc{16}{O} in addition to the fiducial 50\% \nuc{12}{C} and 50\%  \nuc{16}{O}. As expected, the lower the carbon concentration, the larger the critical radii (see fig.~\ref{fig:comp} and tables~\ref{tab:c70}, \ref{tab:c60}, \ref{tab:c40}, \ref{tab:c30}). The important and perhaps surprising result is, that for variations in the carbon fraction within the margin of uncertainty as derived from stellar evolution calculations, the critical radii are more strongly dependent on the \nuc{12}{C} to \nuc{16}{O} ratio than they are on other factors, such as geometry (see section \ref{subsec:geo_res}).

\subsubsection{Admixture of \nuc{4}{He} }
\label{subsec:helium}
Critical radii for initiation decrease as the detonability of the fuel is increased. Similarly, initiation temperatures and densities are much lower if \nuc{4}{He} is added to the fuel (see table~\ref{tab:he4}). The composition of the helium rich fuel considered here consisted of 14\% \nuc{4}{He}, 43\% \nuc{12}{C} and 43\% \nuc{12}{O} by mass, corresponding to a  composition of roughly one \nuc{4}{He} nucleus for every \nuc{12}{C} nucleus. This represents the most favorable condition for rapid energy release via the \mbox{$\nuc{12}{C}  (\alpha,\gamma) \nuc{16}{O}$} channel and can be considered somewhat of a ``sweet spot" of \nuc{4}{He} admixture. For this composition, peak temperatures as low as \mbox{$1.0 \times 10^9$ K} at \mbox{$\rho = 1 \times 10^6$ g cm$^{-3}$} correspond to critical radii on the order of a few tens of km (see table~\ref{tab:he4}).

\subsection{Geometry and curvature}
\label{subsec:geo_res}
The differences in critical radii for the spontaneous initiation of unresolved detonations from a temperature gradient for spherical and planar geometries is moderate. As expected (see section~\ref{ss:geo_curv}), detonations are more reluctant to emerge in spherical geometry.  For the cases considered here, the ratio of the critical sizes in spherical to planar geometry is on the order of a about two or three (see table~\ref{tab:lin_plan}). Curvature effects for gradient initiated detonations thus have a lesser impact on the critical radii compared to the direct initiation problem, where the radii are increased by a factor of around 10 \citep[e.g.][]{he94}. 
We attribute the reduced importance of curvature in this work to the rather different initiation circumstances in the two scenarios.   
The direct initiation problems starts with a very small region of space of extreme over-pressure, a single zone disturbance. 
Curvature effects at early times on the shock of the blast wave are large due to the small radii, i.e. initially $R_c \lesssim l_{det}$, that is, the radius of curvature is smaller than a characteristic length-scale (for example the carbon burning length-scale) of a fully developed detonation. 
In the problem of gradient initiated detonations considered for this work, the shock forms further out, at distances where $R_c >> l_{det}$, leading to a weaker influence of geometrical divergence on the critical initiation parameters.  

\section{Conclusions}
\label{sub:conclusions}
We have presented 1D results of the determination of critical (smallest) spatial scales for the gradient initiation of detonations in white dwarf matter. We have quantified and tabulated the critical sizes for a range of peak and background temperatures, densities, geometry, composition and functional form of the temperature hot-spot. In particular, we have shown that the spatial scales required for the initiation of a detonation is crucially tied to the functional form of the temperature profile.  

An outcome of the work presented in this paper is that the current use of detonation conditions in the literature should be more carefully applied. \citet{fink07}, \citet{roepke07}, \citet{townsley07}, and \citet{jordan08} all rather loosely decide whether or not detonation conditions are reached based solely on the peak temperature reached in the simulation above a certain density. 
The sensitivity of the critical radii on the background temperature and functional form of the profile shows that caution should be exercised when employing such a metric as an authoritative answer to the initiation question.
Absolute limits derived from linear profiles and cold backgrounds such as the minimum temperature or density required for detonation are lowered significantly when taking the possible variation of these parameters into account, possibly turning simulations that were declared ``failures'' into ``successes'' and vice versa.  
It should be emphasized that choosing any such metric relies on the high temperature to occur in unburned fuel, a stipulation which, by the very reactive nature of the fuel, should be at least made plausible to exist.

While this work constitutes an improvement over past efforts, there are several caveats to keep in mind when interpreting the results:
\begin{itemize}
\item The simplified 13-species $\alpha$-chain network is only good within $\sim20\%$ of the energy generation rate of a large network. Since the initiation process is highly nonlinear, a more realistic network may give different results. 
\item The perfectly smooth initial conditions are unrealistic. The effects of noisy initial conditions are unknown, but possibly disturb the formation of detonations.
\item The critical radii presented here result from simulations where the carbon burning is unresolved. While the resolution study (see section~\ref{ss:res_study}) gives some confidence that the corresponding resolved critical radii are merely a factor of two or three larger, there is no guarantee. 
\item Real detonations are three dimensional. 
\end{itemize}
In spite of all these uncertainties, the general trends and conclusions, briefly summarized below, are likely real and should persist.  
\begin{itemize}
\item For the same temperature profile, detonation is more likely at higher density.
\item The dependence on geometry (planar vs spherical) is relatively weak.
\item Composition, especially the presence of \nuc{4}{He}, has a large effect.
\item The background temperature is important (mainly for geometrical reasons) for the critical size determination.
\item The functional form of the temperature profile is a vital component of the initiation process. 
\end{itemize}

It should be clear that, instead of relying on tabulations of detonation conditions such as the ones listed above, it is clearly preferable to resolve the gradients that initiate the detonation in the supernova simulation. 
There is, however, no hope of resolving the carbon burning length scales in a full star simulation in the near future. 
The dependence of the detonation conditions on the functional form of the temperature profile, which is unresolved in all extant multi-dimensional SNe Ia explosion simulations, warrants that definitive and authoritive statements about the failure or success of the initiation of a detonation in such simultions should be issued with caution. The hitherto unknown effects of both non-idealised, noisy initial conditions (for example non monotic temperature profile or non-zero initial velocities) and multi-dimensional instabilities on the critical detonation conditions moreover comprise uncertainties that should be elucidated in the future.  

\acknowledgments
This work is supported in part by the U.S. Department of Energy under contract B523820 to the ASC Alliances Center for Astrophysical Flashes and in part by the National Science Foundation under grant PHY 02-16783 for the Frontier Center "Joint Institute for Nuclear Astrophysics" (JINA). JWT acknowledges support from Argonne National Laboratory, operated under contract No. W-31-109-ENG-38 with the DOE. 

\clearpage

\clearpage

\clearpage

\clearpage

\clearpage

\clearpage

\clearpage

\clearpage

\clearpage

\clearpage

\clearpage

\clearpage
\begin{figure}
\plotone{./f11.eps}
\caption{\label{fig:func24} Critical temperature profiles for \mbox{$T_{max} = 2.4 \times 10^9 $ K} and \mbox{$T_0 = 0.4 \times 10^9$ K.} The density is  \mbox{$10^7$ g cm$^{-3}$. } }
\end{figure}

\clearpage
 \begin{figure}
\plotone{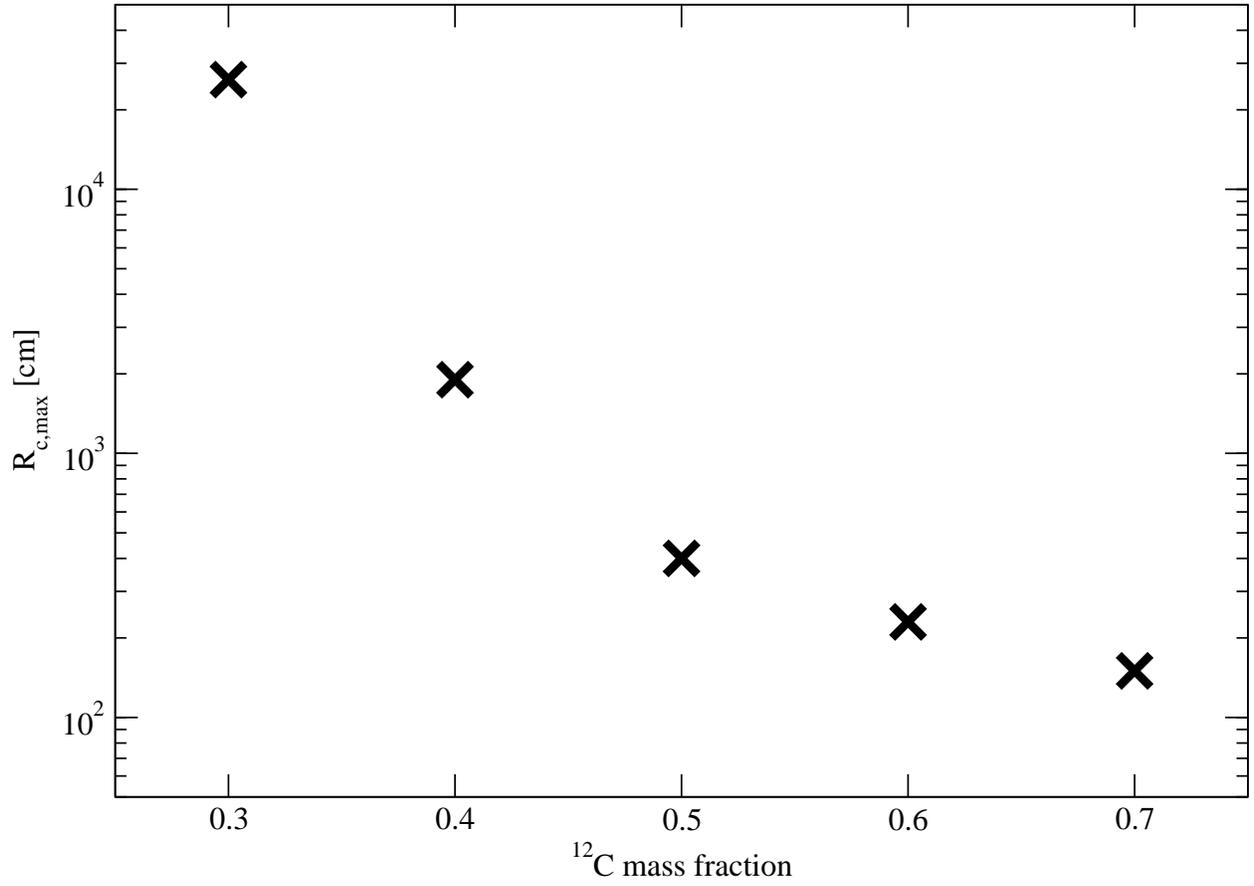}
\caption{\label{fig:comp} Variation of critical radii with composition, all other variables held fixed. Chosen (at random) here $T_{max}=2.4\times10^9$ K, $T_{0}=1.0\times10^9$ K and a linear temperature profile in planar geometry.}
\end{figure}

\clearpage
\begin{deluxetable}{ccrrrcc}
\tabletypesize{\footnotesize}
\tablewidth{0pc}
\tablecaption{Critical Radii for Linear Profile in Planar geometry \label{tab:lin_plan}}
\tablehead{ \colhead{Density} & \colhead{ $T_{max}$}   & \colhead{$T_0$} &
\colhead{$R_{c,min}$} &\colhead{$R_{c,max}$}  &    \colhead{$R^{sph}_{c,max}/R^{plan}_{c,max}$} & \colhead{  $\log (\frac{T_{max}-T_{0}} {R_{c,max}}$) } \\
 \colhead{[ $10^6$ g cm$^{-3}$]} & \colhead{[$10^9$ K]}   & \colhead{[$10^7$ K]} &
\colhead{[$10^2$ cm]} &\colhead{[$10^2$ cm]}  &    \colhead{} &    \colhead{[$\log$(K cm$^{-1}$)] }}
\startdata  
        1            & 2.8        &100.0& 380,000 & 390,000 & -- & 3.66 \\ 
        $\ldots$&$\ldots$&150.0&  24,000 &  25,000 & -- & 4.72 \\ \hline \hline

        10          & 2.8       &   1.0& 1,100 & 1,200 & 1.9 & 6.37 \\ 
        $\ldots$&$\ldots$&  40.0&   860 &   870 & 1.8 & 6.44 \\ 
        $\ldots$&$\ldots$&100.0&   280&   290 & 2.0 &  6.79\\ 
        $\ldots$&$\ldots$&150.0&    62 &    63 & 3.0 & 7.31 \\ \hline
            
        $\ldots$& 2.4    &  1.0& 1,100 & 1,200 & 1.9 &  6.30  \\ 
        $\ldots$&$\ldots$& 40.0&   820 &   830 & 1.8 &  6.38 \\ 
        $\ldots$&$\ldots$&100.0&   390 &   400 & 2.5 & 6.54 \\ 
        $\ldots$&$\ldots$&150.0&   240 &   250 & 2.4 & 6.56 \\ \hline
            
        $\ldots$& 2.0    &  1.0& 6,700 & 6,800 & 2.6 & 5.47 \\ 
        $\ldots$&$\ldots$& 40.0& 5,400 & 5,500 & 2.7 & 5.46 \\ 
        $\ldots$&$\ldots$&100.0& 3,500 & 3,600 & 2.8 & 5.44 \\ 
        $\ldots$&$\ldots$&150.0& 1,400 & 1,500 & 3.1 & 5.52 \\ \hline
            
        $\ldots$& 1.8    &  1.0& 32,000 & 33,000 & -- & 4.73 \\ 
        $\ldots$&$\ldots$& 40.0& 25,000 & 26,000 & -- & 4.73 \\ 
        $\ldots$&$\ldots$&100.0& 11,000 & 12,000 & -- & 4.82 \\ 
        $\ldots$&$\ldots$&150.0&  3,000 &  3,100 & -- & 4.99
        \enddata
  \end{deluxetable}

\begin{deluxetable}{ccc}
\tabletypesize{\footnotesize}\tablewidth{0pc}
\tablecaption{Code to Code Comparison \label{tab:comparison}}
\tablehead{ \colhead{Density} & \colhead{$R_{\mathrm{NW}}$}   & \colhead{$R_{\mathrm{IRS}}$} \\   
\colhead{[$10^7\gcc$]} &\colhead{}  &    \colhead{ }}  
\startdata
1    & 1.0 - 2.0 km & 1.5 - 1.6 km \\
3    & 25 - 50 m    & 38 - 40 m    \\
10   & 1.0 - 2.0 m  & 1.40 - 1.45 m
\enddata
\end{deluxetable}
\begin{deluxetable}{cccrrcc}
\tabletypesize{\footnotesize}
\tablewidth{0pc}
\tablecaption{Resolution Study \label{tab:res_study}}
\tablehead{ \colhead{Density} & \colhead{ $T_{max}$}   & \colhead{$T_0$} &
\colhead{$R_{c,min}$} &\colhead{$R_{c,max}$}  &    \colhead{AMR levels} & \colhead{Resolution}  \\
 \colhead{[$10^6$ g cm$^{-3}$]} & \colhead{[$10^9$ K]}   & \colhead{[$10^7$ K]} &
 \colhead{[$10^2$ cm]} &\colhead{[$10^2$ cm]}  &    \colhead{} &  \colhead{[cm]} }
\startdata  
10      & 2.4            &100.0   &   390 &   400 &  7 &    39.1 \\ 
$\ldots$&$\ldots$&$\ldots$&   770 &   780 &  9 &   19.0 \\ 
$\ldots$&$\ldots$&$\ldots$&   850 &   860 & 11 &  5.2 \\
$\ldots$&$\ldots$&$\ldots$&1,000 &1,100 & 13 & 1.7 \\ 
$\ldots$&$\ldots$&$\ldots$&1,000 &1,100 & 15 & 0.4 \\
$\ldots$&$\ldots$&$\ldots$&1,000 &1,100 & 17 & 0.1 
\enddata
\end{deluxetable}

\begin{deluxetable}{ccrrrc}
\tabletypesize{\footnotesize}
\tablewidth{0pc}
\tablecaption{Critical Radii for Linear Profile in Spherical Geometry\label{tab:lin_sph}}
\tablehead{ \colhead{Density} & \colhead{ $T_{max}$}   & \colhead{$T_0$} &
\colhead{$R_{c,min}$} &\colhead{$R_{c,max}$}  &    \colhead{ $\log (\frac{T_{max}-T_{0}} {R_{c,max}}$) } \\
 \colhead{[$10^6$ g cm$^{-3}$]} & \colhead{[$10^9$ K]}   & \colhead{[$10^7$ K]} &
\colhead{[$10^2$ cm]} &\colhead{[$10^2$ cm]}  &    \colhead{[$\log$(K cm$^{-1}$)] }}
\startdata
   $5.0    $& 3.2    &  1.0& 21,000 & 22,000 & 3.17 \\ 
   $\ldots$&$\ldots$& 40.0& 14,000 & 15,000 & 3.29 \\ 
   $\ldots$&$\ldots$&100.0&  4,200 &  4,300 & 3.71 \\ 
   $\ldots$&$\ldots$&150.0&  1,200 &  1,300 & 4.13 \\ \hline
    
   $\ldots$& 2.8    &  1.0& 20,000 & 21,000 & 3.13 \\ 
   $\ldots$&$\ldots$& 40.0& 14,000 & 15,000 & 3.22 \\ 
   $\ldots$&$\ldots$&100.0&  4,100 &  4,200 & 3.64 \\       
   $\ldots$&$\ldots$&150.0&  1,200 &  1,300 & 4.02 \\ \hline
 
   $\ldots$& 2.4     &  1.0& 21,000 & 22,000 & 3.05 \\ 
   $\ldots$& $\ldots$& 40.0& 13,000 & 14,000 & 3.17 \\ 
   $\ldots$& $\ldots$&100.0&  4,700 &  4,800 & 3.47 \\ 
   $\ldots$& $\ldots$&150.0&  3,100 &  3,200 & 3.46 \\ \hline
      
   $\ldots$& 2.0    &  1.0& 80,000 & 81,000 & 2.39 \\ 
   $\ldots$&$\ldots$& 40.0& 65,000 & 66,000 & 2.39 \\ 
   $\ldots$&$\ldots$&100.0& 40,000 & 41,000 & 2.39 \\    
   $\ldots$&$\ldots$&150.0& 20,000 & 21,000 & 2.39 \\ \hline \hline
       
   $7.0    $& 3.2    &  1.0& 6,900 & 7,000 & 3.66 \\ 
   $\ldots$&$\ldots$& 40.0& 4,700 & 4,800 & 3.77 \\ 
   $\ldots$&$\ldots$&100.0& 1,500 & 1,650 & 4.15 \\
   $\ldots$&$\ldots$&150.0&   470 &   480 & 4.55 \\ \hline 
      
   $\ldots$& 2.8    &  1.0& 6,800 & 6,900 & 3.61 \\ 
   $\ldots$&$\ldots$& 40.0& 4,700 & 4,800 & 3.70 \\ 
   $\ldots$&$\ldots$&100.0& 1,650 & 1,700 & 4.03 \\ 
   $\ldots$&$\ldots$&150.0&   480 &   490 & 4.43 \\ \hline 
  
   $\ldots$& 2.4    &  1.0& 6,800 & 6,900 & 3.54 \\
   $\ldots$&$\ldots$& 40.0& 4,600 & 4,700 & 3.63 \\ 
   $\ldots$&$\ldots$&100.0& 2,100 & 2,200 & 3.81 \\ 
   $\ldots$&$\ldots$&150.0& 1,300 & 1,400 & 3.82 \\ \hline 
      
   $\ldots$& 2.0    &  1.0& 38,000 & 40,000 & 2.71 \\ 
   $\ldots$&$\ldots$& 40.0& 30,000 & 32,000 & 2.71 \\ 
   $\ldots$&$\ldots$&100.0& 19,000 & 20,000 & 2.71 \\ 
   $\ldots$&$\ldots$&150.0&  9,000 & 10,000 & 2.72 \\ \hline \hline

\tablebreak
   
     $10.0  $& 3.2    &  1.0& 2,200 & 2,250 & 4.16 \\ 
  $\ldots$&$\ldots$& 40.0& 1,500 & 1,600 & 4.26 \\ 
  $\ldots$&$\ldots$&100.0&   580 &   590 & 4.58 \\ 
  $\ldots$&$\ldots$&150.0&   180 &   190 & 4.96 \\ \hline
  
  $\ldots$& 2.8    &  1.0& 2,250 & 2,300 & 4.09 \\ 
  $\ldots$&$\ldots$& 40.0& 1,500 & 1,600 & 4.19 \\ 
  $\ldots$&$\ldots$&100.0&   600 &   610 & 4.47 \\ 
  $\ldots$&$\ldots$&150.0&   180 &   190 & 4.85 \\ \hline
            
  $\ldots$& 2.4    &  1.0& 2,200 & 2,250 & 4.03 \\ 
  $\ldots$&$\ldots$& 40.0& 1,400 & 1,500 & 4.14 \\ 
  $\ldots$&$\ldots$&100.0&   900 & 1,000 & 4.17 \\ 
  $\ldots$&$\ldots$&150.0&   590 &   600 & 4.18 \\ \hline
  
  $\ldots$& 2.0    &  1.0& 17,000 & 18,000 & 3.06 \\ 
  $\ldots$&$\ldots$& 40.0& 14,000 & 15,000 & 3.04 \\ 
  $\ldots$&$\ldots$&100.0&  9,000 & 10,000 & 3.02 \\ 
  $\ldots$&$\ldots$&150.0&  4,500 &  4,600 & 3.04 \\ \hline   \hline
    
  $30.0  $& 3.2    &  1.0& 47 & 48 & 5.83 \\ 
  $\ldots$&$\ldots$& 40.0& 39 & 40 & 5.85 \\ 
  $\ldots$&$\ldots$&100.0& 29 & 30 & 5.87 \\
  $\ldots$&$\ldots$&150.0& 14 & 15 & 6.13 \\ \hline

  $\ldots$& 2.8    &  1.0& 54 & 55 & 5.71 \\ 
  $\ldots$&$\ldots$& 40.0& 46 & 47 & 5.71 \\ 
  $\ldots$&$\ldots$&100.0& 34 & 35 & 5.72 \\ 
  $\ldots$&$\ldots$&150.0& 15 & 16 & 5.92 \\ \hline

  $\ldots$& 2.4    &  1.0& 185 & 192 & 5.10 \\ 
  $\ldots$&$\ldots$& 40.0& 150 & 160 & 5.11 \\ 
  $\ldots$&$\ldots$&100.0& 100 & 110 & 5.12 \\ 
  $\ldots$&$\ldots$&150.0&  66 &  67 & 5.13 \\ \hline
      
  $\ldots$& 2.0    &  1.0& 1,700 & 1,800 & 4.06 \\ 
  $\ldots$&$\ldots$& 40.0& 1,400 & 1,500 & 4.04 \\ 
  $\ldots$&$\ldots$&100.0&   900 & 1,000 & 4.02 \\ 
  $\ldots$&$\ldots$&150.0&   480 &   490 & 4.01 
\enddata
\end{deluxetable}

\begin{deluxetable}{ccrrrc}
  \tabletypesize{\footnotesize}
  \tablewidth{0pc}
  \tablecaption{Critical Decay Constants for Exponential Profile in Planar Geometry \label{tab:exp_plan}}
  \tablehead{ \colhead{Density} & \colhead{ $T_{max}$}   & \colhead{$T_0$} &
    \colhead{$R_{c,min}$} &\colhead{$R_{c,max}$}  &    \colhead{$R^{exp}_{c,max}/R^{lin}_{c,max}$} \\
    \colhead{[$10^6$ g cm$^{-3}$]} & \colhead{[$10^9$ K]}   & \colhead{[$10^7$ K]} &
    \colhead{[$10^2$ cm]} &\colhead{[$10^2$ cm]}  &    \colhead{} }
  \startdata  
  10      & 2.4     &  1.0& 890 & 900 &  0.75 \\ 
  $\ldots$& $\ldots$& 40.0& 610 & 620 &  0.75 \\ 
  $\ldots$& $\ldots$&100.0& 340 & 350 & 0.87 \\ 
  $\ldots$& $\ldots$&150.0& 200 & 210 & 0.84 \\ \hline
  
  $\ldots$& 1.8    &  1.0& 28,000 & 29,000 & 0.88 \\ 
  $\ldots$&$\ldots$& 40.0& 21,000 & 22,000 & 0.85 \\ 
  $\ldots$&$\ldots$&100.0&  9,900 & 10,000 & 0.83 \\      
  $\ldots$&$\ldots$&150.0&  2,800 &  2,900 & 0.93 	
  \enddata
\end{deluxetable}

\begin{deluxetable}{ccrrrc}
  \tabletypesize{\footnotesize}
  \tablewidth{0pc}
  \tablecaption{Critical Decay Constants for Gaussian Profile in Planar Geometry \label{tab:gauss_plan}}
  \tablehead{ \colhead{Density} & \colhead{ $T_{max}$}   & \colhead{$T_0$} &
    \colhead{$R_{c,min}$} &\colhead{$R_{c,max}$}  &    \colhead{$R^{gauss}_{c,max}/R^{lin}_{c,max}$} \\
    \colhead{[$10^6$ g cm$^{-3}$]} & \colhead{[$10^9$ K]}   & \colhead{[$10^7$ K]} &
    \colhead{[$10^2$ cm]} &\colhead{[$10^2$ cm]}  &    \colhead{} }
  \startdata  
  1        & 2.4    &100.0& 140,000 & 150,000& -- \\ \hline
  
  10      & 2.4    &  1.0& 1,000 & 1,100& 0.92 \\ 
  $\ldots$&$\ldots$& 40.0&   750 &   760& 0.91 \\ 
  $\ldots$&$\ldots$&100.0&   300 &   310& 0.77 \\ 
  $\ldots$&$\ldots$&150.0&   110 &   120& 0.48 \\ \hline
  
  $\ldots$& 1.8    &  1.0& 9,600 & 9,700 & 0.29 \\ 
  $\ldots$&$\ldots$& 40.0& 7,100 & 7,200 & 0.28 \\ 
  $\ldots$&$\ldots$&100.0& 3,200 & 3,300 &  0.27  \\ 
  $\ldots$&$\ldots$&150.0& 1,100 & 1,200 & 0.39 \\ \hline
  
  $\ldots$& 1.6    &100.0& 8,800 & 8,900 & -- 	
  \enddata
\end{deluxetable}

\begin{deluxetable}{ccrrrc}
  \tabletypesize{\footnotesize}
  \tablewidth{0pc}
  \tablecaption{Critical Decay Constants for g10 in Planar Geometry \label{tab:g10_plan}}
  \tablehead{ \colhead{Density} & \colhead{ $T_{max}$}   & \colhead{$T_0$} &
    \colhead{$R_{c,min}$} &\colhead{$R_{c,max}$}  &    \colhead{$R^{g10}_{c,max}/R^{lin}_{c,max}$} \\
    \colhead{[$10^6$ g cm$^{-3}$]} & \colhead{[$10^9$ K]}   & \colhead{[$10^7$ K]} &
    \colhead{[$10^2$ cm]} &\colhead{[$10^2$ cm]}  &    \colhead{} }
  \startdata  
  10      & 2.4    &  1.0& 1,300 & 1,400 & 1.17 \\ 
  $\ldots$&$\ldots$& 40.0&   960 &   970 & 1.17 \\ 
  $\ldots$&$\ldots$&100.0&   250 &   260 & 0.65 \\ 
  $\ldots$&$\ldots$&150.0&    67 &    68 & 0.27 \\ \hline
  
  $\ldots$& 1.8    &  1.0& 1,500 & 1,600 & 0.048 \\ 
  $\ldots$&$\ldots$& 40.0& 1,000 & 1,100 & 0.042 \\ 
  $\ldots$&$\ldots$&100.0&   750 &   760 & 0.063 \\ 
  $\ldots$&$\ldots$&150.0&   570 &   580 & 0.19 	
  \enddata
\end{deluxetable}

\begin{deluxetable}{ccrrrc}
  \tabletypesize{\footnotesize}
  \tablewidth{0pc}
  \tablecaption{Critical Radii for Linear Profile in Planar Geometry \mbox{70\% \nuc{12}{C} 30\% \nuc{16}{O}} \label{tab:c70}}
  \tablehead{ \colhead{Density} & \colhead{ $T_{max}$}   & \colhead{$T_0$} &
    \colhead{$R_{c,min}$} &\colhead{$R_{c,max}$}  &    \colhead{$R^{70/30}_{c,max}/R^{50/50}_{c,max}$} \\
    \colhead{[$10^6$ g cm$^{-3}$]} & \colhead{[$10^9$ K]}   & \colhead{[$10^7$ K]} &
    \colhead{[$10^2$ cm]} &\colhead{[$10^2$ cm]}  &    \colhead{} }
  \startdata  
  10      & 2.4    &  1.0& 240 & 250 & 0.21 \\ 
  $\ldots$&$\ldots$& 40.0& 210 & 220 & 0.26 \\ 
  $\ldots$&$\ldots$&100.0& 140 & 150 & 0.30 \\ 
  $\ldots$&$\ldots$&150.0&  94 &  95 & 0.38
  \enddata
\end{deluxetable}

\begin{deluxetable}{ccrrrc}
  \tabletypesize{\footnotesize}
  \tablewidth{0pc}
  \tablecaption{Critical Radii for Linear Profile in Planar Geometry \mbox{60\% \nuc{12}{C} 40\% \nuc{16}{O}} \label{tab:c60}}
  \tablehead{ \colhead{Density} & \colhead{ $T_{max}$}   & \colhead{$T_0$} &
    \colhead{$R_{c,min}$} &\colhead{$R_{c,max}$}  &    \colhead{$R^{60/40}_{c,max}/R^{50/50}_{c,max}$} \\
    \colhead{[$10^6$ g cm$^{-3}$]} & \colhead{[$10^9$ K]}   & \colhead{[$10^7$ K]} &
    \colhead{[$10^2$ cm]} &\colhead{[$10^2$ cm]}  &    \colhead{} }
  \startdata  
  10      & 2.4    &  1.0& 370 & 380 & 0.32 \\ 
  $\ldots$&$\ldots$& 40.0& 320 & 330 & 0.40 \\ 
  $\ldots$&$\ldots$&100.0& 220 & 230 & 0.57 \\ 
  $\ldots$&$\ldots$&150.0& 140 & 150 & 0.60
  \enddata
\end{deluxetable}

\begin{deluxetable}{ccrrrc}
  \tabletypesize{\footnotesize}
  \tablewidth{0pc}
  \tablecaption{Critical Radii for Linear Profile in Planar Geometry \mbox{40\% \nuc{12}{C} 60\% \nuc{16}{O}} \label{tab:c40}}
  \tablehead{ \colhead{Density} & \colhead{ $T_{max}$}   & \colhead{$T_0$} &
    \colhead{$R_{c,min}$} &\colhead{$R_{c,max}$}  &    \colhead{$R^{40/60}_{c,max}/R^{50/50}_{c,max}$} \\
    \colhead{[$10^6$ g cm$^{-3}$]} & \colhead{[$10^9$ K]}   & \colhead{[$10^7$ K]} &
    \colhead{[$10^2$ cm]} &\colhead{[$10^2$ cm]}  &    \colhead{} }
  \startdata  
  10      & 2.4    &  1.0& 7,000 & 7,100& 5.92 \\ 
  $\ldots$&$\ldots$& 40.0& 5,500 & 5,600& 6.75 \\ 
  $\ldots$&$\ldots$&100.0& 1,800 & 1,900& 4.75 \\ 
  $\ldots$&$\ldots$&150.0&   520 &   530& 2.12 
  \enddata
\end{deluxetable}

\begin{deluxetable}{ccrrrc}
  \tabletypesize{\footnotesize}
  \tablewidth{0pc}
  \tablecaption{Critical Radii for Linear Profile in Planar Geometry \mbox{30\% \nuc{12}{C} 70\% \nuc{16}{O}} \label{tab:c30}}
  \tablehead{ \colhead{Density} & \colhead{ $T_{max}$}   & \colhead{$T_0$} &
    \colhead{$R_{c,min}$} &\colhead{$R_{c,max}$}  &    \colhead{$R^{30/70}_{c,max}/R^{50/50}_{c,max}$} \\
    \colhead{[$10^6$ g cm$^{-3}$]} & \colhead{[$10^9$ K]}   & \colhead{[$10^7$ K]} &
    \colhead{[$10^2$ cm]} &\colhead{[$10^2$ cm]}  &    \colhead{} }
  \startdata  
  10      & 2.4    &  1.0& 50,000 & 51,000 &  42.50 \\ 
  $\ldots$&$\ldots$& 40.0& 46,000 & 47,000 & 56.63 \\ 
  $\ldots$&$\ldots$&100.0& 25,000 & 26,000 & 65.00 \\ 
  $\ldots$&$\ldots$&150.0&  2,300 & 2,400  & 9.60
  \enddata
\end{deluxetable}

\begin{deluxetable}{ccrrrc}
  \tabletypesize{\footnotesize}
  \tablewidth{0pc}
  \tablecaption{Critical Radii for 14\% \nuc{4}{He} , 33\% \nuc{12}{C}, and 33\% \nuc{16}{O} \label{tab:he4} }
  \tablehead{ \colhead{Density} & \colhead{ $T_{max}$}   & \colhead{$T_0$} &
    \colhead{$R_{c,min}$} &\colhead{$R_{c,max}$}  &    \colhead{ $\log (\frac{T_{max}-T_{0}} {\bar{R}_{c,max}}$) } \\
    \colhead{[$10^6$ g cm$^{-3}$]} & \colhead{[$10^9$ K]}   & \colhead{[$10^7$ K]} &
    \colhead{[$10^2$ cm]} &\colhead{[$10^2$ cm]}  &    \colhead{[$\log$(K cm$^{-1}$)] }}
  \startdata
  $ 1 $     & $2.0$  & $ 1.0 $ & 7,000 & 7,500 & 5.42 \\ 
  $ \dots $ & $\dots $  & $ 40.0 $ & 4,500 & 5,000 & 5.38 \\  \hline       
  $\ldots $ & $1.6$  & $ 1.0 $ & 8,000 & 8,500 & 5.27 \\    
  $\ldots $ & $\dots $  & $ 40.0 $ & 5,000 & 5,500 & 5.34 \\  \hline    
  $\ldots $ & $1.2$  & $ 1.0 $ & 14,000 & 15,000 &4.90 \\    
  $\ldots $ & $\dots $  & $ 40.0 $ & 9,500 & 10,000 &4.90 \\  \hline     
  $\ldots $ & $1.0$  & $ 1.0 $ & 32,000 & 35,000 &4.45 \\    
  $\ldots $ & $\dots $  & $ 40.0 $ & 19,000 & 20,000 & 4.48 \\ \hline \hline
  
  $ 10 $    & $2.0$  & $ 1.0 $ & 58 & 61 & 7.51 \\ 
  $\ldots $ & $\dots $  & $ 40.0 $ & 45 & 50 & 7.51 \\  \hline         
  $\ldots $ & $1.6$  & $ 1.0 $ & 100 & 110 & 7.16 \\    
  $\ldots $ & $\dots $  & $ 40.0 $ & 80 & 85 & 7.15 \\  \hline      
  $\ldots $ & $1.2$  & $ 1.0 $ & 275 & 288 & 6.62  \\    
  $\ldots $ & $\dots $  & $ 40.0 $ & 188 & 200 & 6.60  \\  \hline 
  $\ldots $ & $1.0$  & $ 1.0 $ & 600 & 650 & 6.18 \\    
  $\ldots $ & $\dots $  & $ 40.0 $ & 350 & 380 & 6.20 
  \enddata
\end{deluxetable}

\end{document}